\documentclass[aps,pre,twocolumn]{revtex4}
\usepackage{braket}
\usepackage[pdftex]{graphicx}

\begin{document}
\title{A unified statistical model of protein multiple sequence alignment integrating direct coupling and insertions}
\author{Akira R. Kinjo}
\email{akinjo@protein.osaka-u.ac.jp}
\affiliation{Institute for Protein Research, Osaka University, 
3-2 Yamadaoka, Suita, Osaka, 565-0871, Japan}
%\date{$ $Date$ $}
\date{\today}

\begin{abstract}
The multiple sequence alignment (MSA) of a protein family provides a wealth of information in terms of the conservation pattern of amino acid  residues not only at each alignment site but also between distant sites.  In order to statistically model the MSA incorporating both short-range and  long-range correlations as well as insertions, I have derived a lattice  gas model of the MSA based on the principle of maximum entropy.  The partition function, obtained by the transfer matrix method with a  mean-field approximation, accounts for all possible alignments with all  possible sequences. The model parameters for short-range  and long-range interactions were determined by a self-consistent  condition and by a  Gaussian approximation, respectively.  Using this model with and without long-range interactions,  I analyzed the globin and V-set domains by increasing the ``temperature''  and by ``mutating'' a site.  The correlations between residue conservation and various measures of  the system's stability indicate that the long-range interactions  make the conservation pattern more specific to the structure, and  increasingly stabilize better conserved residues.
\end{abstract}


\maketitle
\section{Introduction}
\label{sec:Introduction}
A multiple sequence alignment (MSA) of a family of proteins provides us with valuable information to characterize the protein family in terms of patterns of amino acid residues at alignment sites~\cite{DurbinETAL}. The usefulness of analyzing the residue compositions in the MSA has led to the development of a class of sequence profile methods~\cite{Taylor1986,GribskovETAL1987,DurbinETAL} such as PSI-BLAST~\cite{AltschulETAL1997} and profile hidden Markov models (HMM)~\cite{KroghETAL1994}, which can be used to detect distantly related proteins, to obtain high-quality alignments, and to improve structure prediction~\cite{Rost2003} as well as to characterize functional and structural roles of the conservation pattern~\cite{KinjoANDNakamura2008}. In the sequence profile methods, it is assumed that the residue composition of each site is independent of other sites. With this crude assumption, the conservation of residues are explained in terms of their functional and structural roles. However, to further understand
the mechanism of these roles in the context of protein sequences,
one needs to drop the assumption of site independence.
In fact, there seems to be no way for a residue to ``know'' that it is in
a particular position in the sequence to play a particular functional
or structural role other than by its interactions with other residues in
the sequence (or with other molecules in the biological system). Therefore, to
understand what makes particular residues important at each site,
one needs to study the correlations between different sites. 

Correlations between distant sites in a MSA can be quantified by identifying
correlated substitutions. They have been exploited to gain further insights
of structures and functions of proteins~\cite{Toh2004,deJuanETAL2013,TaylorETAL2013}. However, the apparent correlations observed in a MSA are only a result of intricate interactions between residues as in the underlying native structures of proteins. Recently, there have been a number of successful attempts to extract direct correlations~\cite{deJuanETAL2013,TaylorETAL2013} which are in fact found to be in excellent agreement with the residue-residue contacts in native structures~\cite{MorcosETAL2011,JonesETAL2012,Miyazawa2013}
to the extent that the three-dimensional structures can be actually (re)constructed~\cite{MarksETAL2011,TaylorETAL2012}.

One drawback of the direct-coupling analysis (as well as other direct correlation methods) is that it takes into account only those alignment sites that are well aligned (the ``core'' sites), and ignores insertions. The primary difficulty in the treatment of insertion is that they are of variable lengths, which makes the system size variable and hence greatly complicates the problem. When one is interested in some universal properties of a protein family such as their approximate three-dimensional fold, insertions may be irrelevant. However, when one is interested in a particular member of the family, the existence of some insertions may be important. In fact, insertions, which may be regarded as ``embellishments'' to a conserved structural core, are deemed to be an effective strategy for proteins to diversify and specialize their functions~\cite{DessaillyETAL2009}. Some insertions are also known to play critical roles in protein oligomerization~\cite{HashimotoETAL2010,NishiETAL2011}. Of more fundamental concern is that ignoring insertions in a MSA means ignoring the polypeptide chain structure, which implies theoretical as well as practical consequences. Theoretically,
it is questionable to ignore such a strong interaction as the peptide bond in order to accurately describe the sequence and structure of proteins.
Practically, in order to identify new members of a family by aligning their sequences to some MSA-derived model incorporating direct correlations, a consistent treatment of polypeptide sequences is necessary.

In this paper, I present a new statistical model of the MSA that incorporates both direct correlations and insertions. The main objective of this model is incorporation of long-range correlations into multiple-sequence alignment, rather than improving contact prediction by incorporating insertions. As will be apparent from the formulation, this model is a generalization of the direct-coupling analysis that is based on the principle of maximum entropy~\cite{LapedesETAL1999,MorcosETAL2011}. This model can be regarded as a finite, quasi-one-dimensional, multicomponent, and heterogeneous lattice gas model where the ``particles'' are amino acid residues.
In the following, the ``lattice gas model'' refers to this model. The lattice system consists of two kinds of lattice sites, corresponding to the core (matching or deletion) or the insert, that are connected in a similar, but distinctively different, manner as in the profile HMM model. While long-range interactions are treated by using a mean-field approximation, short-range interactions are treated rigorously so that the partition function is obtained analytically by a transfer matrix method. One notable feature of this model is that its partition function literally accounts for all the possible alignments with all the possible protein sequences, including infinitely long ones.
Based on this model,
 various virtual experiments can be performed by changing the ``temperature'' of the system or by manipulating the ``chemical potentials'' associated with the particles (residues) at each site. 

The paper is organized as follows.
In Section \ref{sec:Theory},  some basic quantities are defined and the lattice gas model of the multiple sequence alignment is formulated. Section \ref{sec:Materials} provides the details of numerical methods and data preparation. Section \ref{sec:Results} gives the results of virtual experiments by increasing the temperature or by introducing alanine point mutants. In Section \ref{sec:Discussion}, limitations, implications as well as possible extensions of
the present model are discussed.
\section{Theory}
\label{sec:Theory}
\begin{figure}
\begin{center}
\includegraphics[width=8cm]{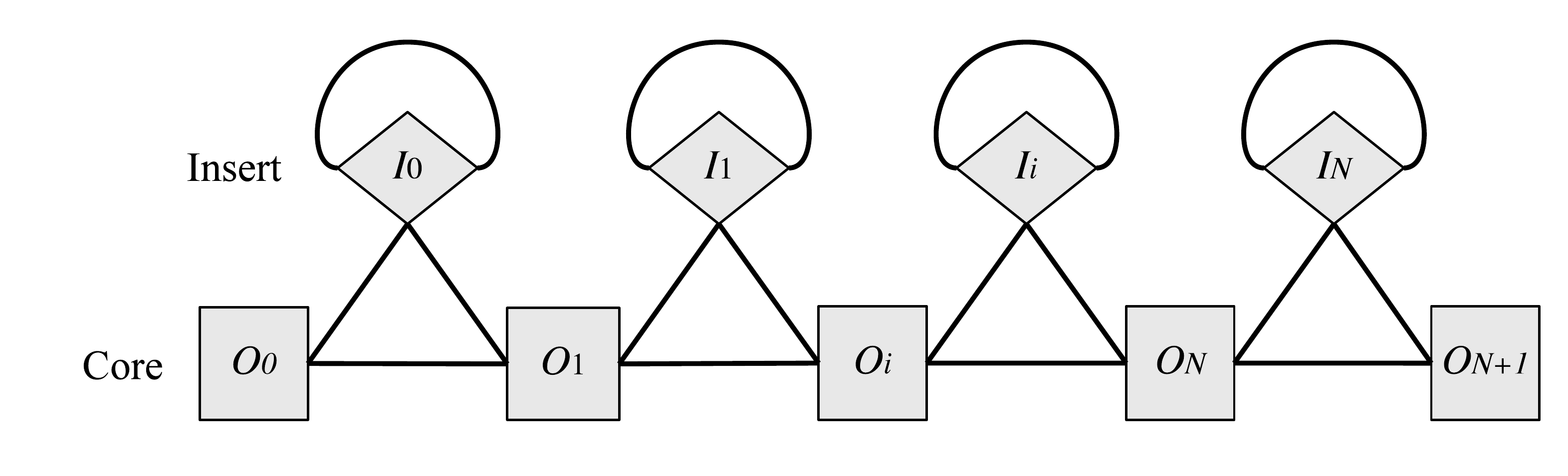}
\end{center}
\caption{\label{fig:model}
The lattice structure of the model. The squares marked with $O_i  (i = 0,\cdots, N+1)$ correspond to core (matching/deletion) sites, the diamonds marked with $I_i (i = 0, \cdots, N)$ correspond to insert sites. The edges between sites indicates bonded interactions. See Figure \ref{fig:alignment} for concrete examples.}
\end{figure}

\subsection{Representing multiple sequence alignment as lattice gas system}
A MSA may be regarded as a matrix of symbols in which each row is a protein sequence possibly with gaps and each column is an alignment site. 
Some columns may contain few gaps so the residues in such positions may be 
relatively important for the protein family. Here, I informally define 
a ``core'' (matching/deletion) site as an alignment site
which are relatively well aligned.
The remaining sites are defined to be insert sites. Core sites are 
ordered from the N-terminal to the C-terminal, and denoted 
as $O_1, O_2, \cdots, O_N$ with $N$ being the number of core sites. 
For convenience, the terminal core sites $O_0$ and $O_{N+1}$ are appended to indicate the start and end of the alignment, as in the profile HMM~\cite{DurbinETAL}. To each core site, either one of 20 amino acid residues or a gap (deletion) may be assigned, and the latter is treated as the 21-st type of residue. An insert site between two core sites $O_i$ and $O_{i+1}$ is denoted as $I_i$.
All the gap symbols ignored at an insert site. In the following, the (ordered) sets of core and insert sites are denoted as $\mathcal{O}=\{O_0\cdots,O_{N+1}\}$ and $\mathcal{I} = \{I_0,\cdots,I_N\}$, respectively, and their union as $\mathcal{S} = \mathcal{O} \cup \mathcal{I}$. In addition, let us define a set of amino acid residues allowed for an insert site $I_i$ as $\mathcal{A}_{I_i} = \{\mathtt{A},\cdots,\mathtt{Y}\}$ (20 amino acid residue types), and that for a core site $O_i$ as $\mathcal{A}_{O_i} = \{\mathtt{A},\cdots,\mathtt{Y},\mathtt{-}\}$ (20 amino acid residues and deletion) for $i = 1,\cdots,N$ and $\mathcal{A}_{O_0} = \mathcal{A}_{O_{N+1}} = \{\mathtt{-}\}$ (deletion only) for the terminal sites. 

For one protein sequence in the MSA, at most one residue may correspond to each core site $O_i$ whereas any number of residues may correspond to an insert site $I_i$. In this sense, residues behave like fermions on core sites and like bosons on insert sites. The set of core and insert sites comprise a quasi-one-dimensional lattice structure as shown in Figure \ref{fig:model}. 
In this lattice structure, two sites are connected if two consecutive residues  in a protein sequence (possibly including gap symbols) can be assigned. If two sites are directly connected, they are defined to be a bonded or short-range pair. 
The self-connecting loop in each insert site indicates that it makes a bonded pair with itself. Thus, an insertion may be indefinitely long, manifesting its boson-like character. 

\begin{figure}
\begin{center}
\includegraphics[width=8cm]{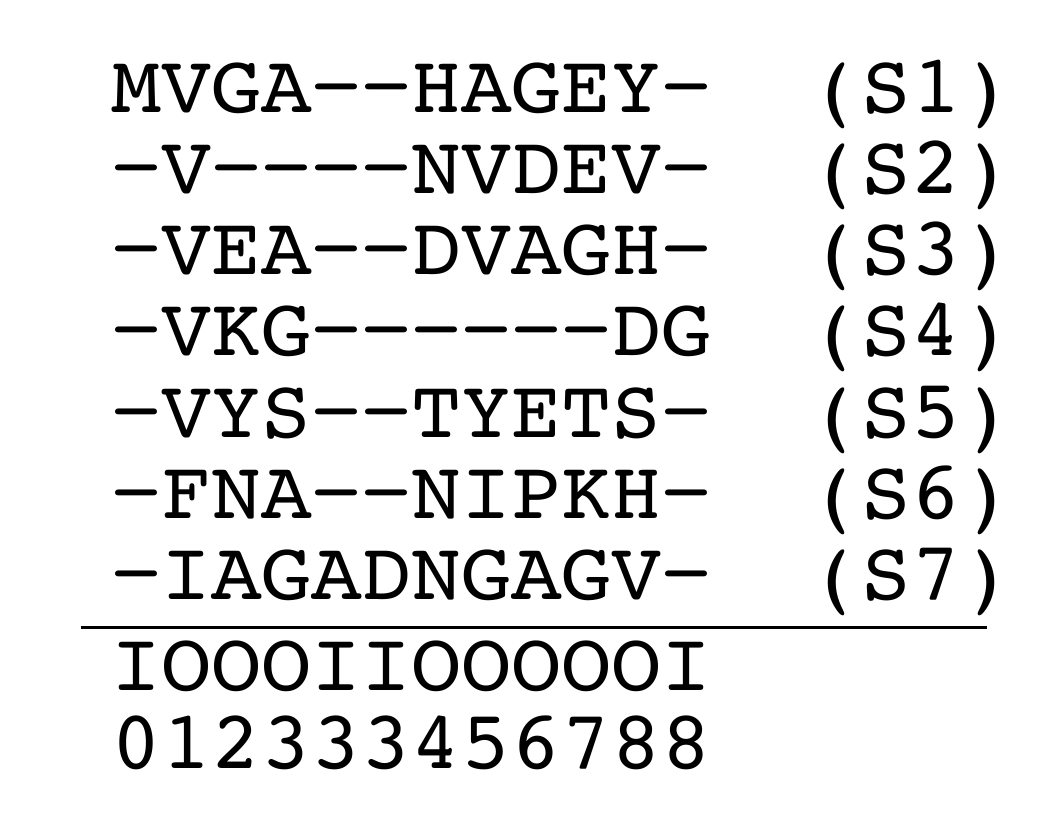}
\end{center}
\caption{\label{fig:alignment}
Example of a multiple sequence alignment (based on~\cite{DurbinETAL}). Each row corresponds to a protein sequence (S1,$\cdots$, S7) and each column to an alignment site. Below the horizontal line, each alignment site is annotated as to whether it corresponds to a core (matching or deletion) site (``O'') or an insert site (``I''). Indicated below these ``O''/ ``I'' symbols are the position of lattice sites. (c.f. Figure \ref{fig:model}.) The size of the lattice model based on this MSA is $N=8$. Insert sites other than $I_0$, $I_3$ and $I_8$ are not explicit in this MSA.
For example, the alignment of the sequence S2 in this figure is represented as $\mathbf{X}^{S2} = X_0\cdots X_9 =$
$(O_0,\mathtt{-})(O_1,\mathtt{V})$
$(O_2,\mathtt{-})$
$(O_3,\mathtt{-})$
$(O_4,\mathtt{N})$
$(O_5,\mathtt{V})$
$(O_6,\mathtt{D})$
$(O_7,\mathtt{E})$
$(O_8,\mathtt{V})$
$(O_9,\mathtt{-})$ where the first and last pairs represent the start and end of  the alignment, respectively.
As another example, the alignment of sequence S7 is 
$\mathbf{X}^{S7} = X_0\cdots X_{11} =$
$(O_0,\mathtt{-})$
$(O_1,\mathtt{I})$
$ (O_2,\mathtt{A})$
$ (O_3,\mathtt{G})$
$ (I_3,\mathtt{A})$
$ (I_3,\mathtt{D})$
$ (O_4,\mathtt{N})$
$ (O_5,\mathtt{G})$
$ (O_6,\mathtt{A})$
$ (O_7,\mathtt{G})$
$ (O_8,\mathtt{V})$
$(O_9,\mathtt{-})$.
}
\end{figure}

Based on this lattice system, an alignment $\mathbf{X}$ of a particular protein sequence $\mathbf{a} = a_1a_2\cdots a_L$ in the MSA may be represented as a sequence of length $L_{\mathbf{X}}$ consisting of ordered pairs of a lattice site and a residue of $\mathbf{a}$: $\mathbf{X} = X_0X_1\cdots X_{L_{\mathbf{X}}}X_{L_{\mathbf{X}}+1}$ (``matchings'' to the terminal sites are also included). Here, each $X_k = (S,a)$ with $S\in \mathcal{S}$ and $a \in \mathcal{A}_S$. A whole MSA consisting of $M$ sequences is a set of such aligned sequences: $\{\mathbf{X}^{t}\}_{t=1,\cdots,M}$. Figure \ref{fig:alignment} shows some concrete examples of this representation of alignment.

\subsection{Variables to characterize alignments}
Using the above representation, let us define some quantities that
characterize an alignment in a given MSA.
For a given lattice model and its alignment $\mathbf{X}$ with a protein sequence, the number of the residue type $a \in \mathcal{A}_S$ at the lattice site $S \in \mathcal{S}$ is defined as
\begin{equation}
  \label{eq:sscount}
n_{S}(a|\mathbf{X}) = \sum_{k=0}^{L_{\mathbf{X}}}\delta_{(S,a),X_k}.
\end{equation}
This quantity is referred to as the single-site count.
Similarly, the number of a pair of residue types $a \in \mathcal{A}_S$ and $b\in \mathcal{A}_{S'}$ on a bonded pair of lattice sites $S$ and $S'$ occupied by two consecutive alignment sites
is defined as
\begin{equation}
  \label{eq:bpcount}
n_{SS'}^b(a,b|\mathbf{X}) = \sum_{k=0}^{L_\mathbf{X}} \delta_{(S,a),X_{k}} \delta_{(S',b),X_{k+1}},
\end{equation}
which is referred to as the bonded pair count.
The single-site counts and bonded pair counts are the two fundamental stochastic variables in the present theory. For later convenience, let us define the non-bonded pair counts as
\begin{equation}
  \label{eq:nbcount}
n_{SS'}^{nb}(a,b|\mathbf{X}) = n_{S}(a|\mathbf{X})n_{S'}(b|\mathbf{X})
\end{equation}
for $S, S' \in \mathcal{S}$.
Note that the non-bonded pair counts may be defined for residues residing on neighboring lattice sites as well as on the same ($S=S'$) site. The terms ``bonded'' and ``non-bonded'' here are meant to describe the connectivity along the polypeptide sequence rather than that along the lattice system (A pair of residues in neighboring lattice sites may be either bonded or non-bonded depending on the given alignment).
From these definitions, several relations follow. First, by the fermion-like character of the core site, we have for each $O_i \in \mathcal{O}$
\begin{equation}
  \label{eq:rel1}
  \sum_{a\in \mathcal{A}_{O_i}}n_{O_i}(a|\mathbf{X}) = 1.
\end{equation}
Between bonded pair counts and single-site count, we have
\begin{eqnarray}
  \sum_{b\in\mathcal{A}_{O_{i+1}}}n_{S,O_{i+1}}^{b}(a,b|\mathbf{X})
  + \sum_{b\in\mathcal{A}_{I_{i}}}n_{S,I_{i}}^{b}(a,b|\mathbf{X}) & = & n_{S}(a|\mathbf{X}),\nonumber\\
\label{eq:relb1}  &&\\
  \sum_{a\in\mathcal{A}_{O_{i}}}n_{O_{i},S'}^{b}(a,b|\mathbf{X})
  + \sum_{a\in\mathcal{A}_{I_{i}}}n_{I_{i},S'}^{b}(a,b|\mathbf{X}) & = & n_{S'}(b|\mathbf{X})\nonumber\\
\label{eq:relb2}  &&
\end{eqnarray}
where $S = O_i, I_i$ and $S' = O_{i+1}, I_{i}$.
Lastly, between non-bonded pair counts and single-site count, we have
\begin{eqnarray}
  \label{eq:relnb1}
  \sum_{b\in\mathcal{A}_{O_j}}n_{S,O_j}^{nb}(a,b|\mathbf{X}) &=& n_{S}(a|\mathbf{X}),\\
  \label{eq:relnb2}
    \sum_{a\in\mathcal{A}_{O_i}}n_{O_i,S'}^{nb}(a,b|\mathbf{X}) &=& n_{S'}(b|\mathbf{X})
\end{eqnarray}
where $S,S'\in\mathcal{S}$.

\subsection{Probability distribution of alignments}
I would like to statistically characterize the given MSA in terms of the above quantities.
To do so, suppose that the probability $P(\mathbf{X})$ of an alignment $\mathbf{X}$ is known for the lattice model. Then, the expectation values of these numbers are defined as follows:
\begin{eqnarray}
{n}_{S}(a) &=& \sum_{\mathbf{X}}P(\mathbf{X})n_{S}(a|\mathbf{X}),\label{eq:n}\\
{n}_{SS'}^b(a,b) &=& \sum_{\mathbf{X}}P(\mathbf{X})
n_{SS'}^b(a,b|\mathbf{X}),
\label{eq:ns}\\
n_{SS'}^{nb}(a,b) &=& \sum_{\mathbf{X}}P(\mathbf{X})n_{SS'}^{nb}(a,b|\mathbf{X})
\label{eq:nl}
\end{eqnarray}
which are referred to as single-site (number) densities, bonded pair (number) densities, and non-bonded pair (number) densities, respectively.
These number densities naturally satisfy the relations analogous to Eqs. (\ref{eq:rel1})-(\ref{eq:relnb2}).

To determine the form of $P(\mathbf{X})$, the principle of maximum entropy is employed with the constraints that the densities are equal to those observed in the given MSA. The entropy is given as
\begin{equation}
  \label{eq:entropy}
  S  = -\sum_{\mathbf{X}}P(\mathbf{X})\ln P(\mathbf{X}).
\end{equation}
Let us denote the densities estimated from the given MSA as $\bar{n}_{S}(a)$, $\bar{n}_{SS'}^b(a,b)$, and $\bar{n}_{SS'}^{nb}(a,b)$ (see Section \ref{sec:Materials} for the method to obtain these quantities). 
The following Lagrangian, consisting of the entropy (Eq. \ref{eq:entropy}) and the constraints for the densities, is maximized:
\begin{eqnarray}
  \label{eq:lagrangian}
  \mathcal{L} & = &
-T\sum_{\mathbf{X}}P(\mathbf{X})\ln P(\mathbf{X})\nonumber\\
&&
+{\sum_{(S,S')}^{\mathrm{b.p.}}}\sum_{a,b}J_{SS'}(a,b)[
{n}_{SS'}^{b}(a,b) - \bar{n}_{SS'}^{b}(a,b)]\nonumber\\
&&+\frac{1}{2}\sum_{S,S'}\sum_{a,b}K_{SS'}(a,b)[
{n}_{SS'}^{nb}(a,b) - \bar{n}_{SS'}^{nb}(a,b)]\nonumber\\
&&+ \sum_{S,a}\mu_{S}(a)[{n}_{S}(a) - \bar{n}_{S}(a)]
\end{eqnarray}
where $\mu_{S}(a)$, $J_{SS'}(a,b)$ and $K_{SS'}(a,b)$ are undetermined multipliers, and the summation ${\sum_{(S,S')}^{\mathrm{b.p.}}}$ is over bonded pairs. We have also introduced the ``temperature'' parameter $T$. Solving $\delta \mathcal{L}/\delta P(\mathbf{X}) = 0$ leads to the Boltzmann distribution:
\begin{equation}
  \label{eq:boltzmann}
  P(\mathbf{X}) = \frac{\exp[-E(\mathbf{X})/T]}{\Xi},
\end{equation}
where $\Xi$ is the normalization constant or the partition function defined by
\begin{equation}
\Xi = \sum_{\mathbf{X}}\exp[-E(\mathbf{X})/T],\label{eq:partition-function}
\end{equation}
and $E(\mathbf{X})$ is the ``energy'' of the system given as
\begin{eqnarray}
  \label{eq:energy}
  E(\mathbf{X}) &=& 
-{\sum_{(S,S')}^{\mathrm{b.p.}}}\sum_{a,b}J_{SS'}(a,b)n_{SS'}^b(a,b|\mathbf{X})\nonumber\\
&& -\frac{1}{2}\sum_{S,S'}\sum_{a,b}K_{SS'}(a,b)n_{S}(a|\mathbf{X})n_{S'}(b|\mathbf{X})\nonumber\\
&& -\sum_{S,a}\mu_{S}(a)n_{S}(a|\mathbf{X}).
\end{eqnarray}
From this expression of the energy function, we can interpret $\mu_{S}(a)$ as the chemical potential imposed on the particle (amino acid residue) $a$ at the site $S$, and $J$ and $K$ as bonded and non-bonded coupling parameters, respectively.
The problem of obtaining the probability distribution $P(\mathbf{X})$ is thus reduced to computing the partition function $\Xi$. In the following, the non-bonded interactions are considered only between core sites (i.e., core-insert and insert-insert pairs are discarded) for a technical reason (see the subsection ``Determining the $K$ matrix'' below).

\subsection{Partition function}
In this subsection, I assume that the parameters $\mu, J$ and $K$ are fixed.
To treat the long-range interactions, a mean-field approximation is applied. Then, the partition function can be computed by a transfer matrix method. Let us define the mean field $K_{S}(a)$ acting on the residue type $a$ on the site $S$:
\begin{equation}
  \label{eq:mean-field}
  \tilde{K}_{S}(a) = \sum_{S',b}K_{SS'}(a,b)[{n_{S'}(b)} - \bar{n}_{S'}(b)]
\end{equation}
where $\bar{n}_{S'}(b)$ is subtracted for convenience, but this does not essentially change the system's behavior (it simply shifts the chemical potential $\mu_{S}(a)$ which can be compensated by $J$; see Eq. \ref{eq:tmat} and Section \ref{sec:gauge}).
Next, let us define the transfer matrices between a bonded pair of sites $S = O_i, I_i$ and $S' = O_{i+1}, I_i$ as
\begin{eqnarray}
  T_{SS'}(a,b)& =& \exp[\{J_{SS'}(a,b) + \mu_{S'}(b) + \tilde{K}_{S'}(b)\}/T].\nonumber\\
&&  \label{eq:tmat}
\end{eqnarray}

To alleviate the expressions for the partial partition functions, a bracket notation is introduced. First, define a set of standard basis vectors: $\bra{a}$ and $\ket{a}$ corresponding to each residue type $a$ on each site. These vectors satisfy the following orthonormal properties:
\begin{eqnarray}
\braket{a|b} &=& \delta_{a,b},\\
\label{eq:complete}
\sum_{a\in\mathcal{A}_S} \ket{a}\bra{a} &=& \mathbf{I}_{|\mathcal{A}_S|} \mbox{(identity matrix)}
\end{eqnarray}
where $\mathbf{I}_{|\mathcal{A}_S|}$ is the ${|\mathcal{A}_S|}$-dimensional identity matrix.
For each site $i$, I define the partial partition functions $\bra{O_i}$ and $\bra{I_i}$
that count the statistical weight of all possible alignments starting from the start site $O_0$ and terminating at $O_i$ and $I_i$, respectively. Similarly, partial partition functions $\ket{O_i}$ and $\ket{I_i}$ account for all possible alignments ``starting'' from the end site $O_{N+1}$ and ``terminating'' at $O_i$ and $I_i$.
Any (complete) alignment starts at the start site $O_0$ and ends at the end site $O_{N+1}$, and these sites are formally treated as ``deletion (\texttt{-}).'' Therefore, the boundary conditions are given as
\begin{eqnarray}
\bra{O_0} &=& \bra{\mathtt{-}} = (0,\cdots,0,1),\\
\ket{O_{N+1}} &=& \ket{\mathtt{-}} = (0,\cdots,0,1)^t.
\end{eqnarray}
Based on this setting, the recursion formulae for partial partition functions are given as
\begin{eqnarray}
  \label{eq:forwardO}
  \bra{O_{i+1}} &=& \bra{O_{i}}T_{O_{i}O_{i+1}} + \bra{I_{i}}T_{I_{i}O_{i+1}},\\
 \label{eq:forwardI} \bra{I_i} &=& \bra{O_{i}}T_{O_{i}I_{i}}
  + \bra{I_{i}}T_{I_{i}I_{i}}
\end{eqnarray}
in the forward (N- to C-terminal) direction, and
\begin{eqnarray}
\label{eq:backwardO}
\ket{O_i} &= & T_{O_iO_{i+1}}\ket{O_{i+1}} + T_{O_iI_i}\ket{I_i},\\
\label{eq:backwardI}
\ket{I_i} &= & T_{I_iO_{i+1}}\ket{O_{i+1}} + T_{I_iI_i}\ket{I_i}
\end{eqnarray}
in the backward (C- to N-terminal) direction. Here, each transfer matrix $T_{SS'}$ is viewed as a $|\mathcal{A}_{S}|\times |\mathcal{A}_{S'}|$ matrix
with $\braket{a|T_{SS'}|b} = T_{SS'}(a,b)$.
By expanding Eq. (\ref{eq:forwardI}), we have
\begin{eqnarray}
\bra{I_i} &=& \bra{O_i}T_{O_iI_i}\left(\mathbf{I} + {T_{I_iI_i}} + {T_{I_iI_i}^2} + \cdots\right)\\
&=& \bra{O_i}T_{O_iI_i}(\mathbf{I} - T_{I_iI_i})^{-1}
\label{eq:ITII}
\end{eqnarray}
where $\mathbf{I} = \mathbf{I}_{20}$ (the 20-dimensional identity matrix). Similarly, we have
\begin{eqnarray}
  \ket{I_i} &=& (\mathbf{I} - T_{I_iI_i})^{-1}T_{I_iO_{i+1}}\ket{O_{i+1}}.
  \label{eq:ITII2}
\end{eqnarray}
Thus, $\bra{I_i}$ and $\ket{I_i}$ indeed include contributions from infinitely long insertions.
The inverse matrix $(\mathbf{I} - T_{I_iI_i})^{-1}$ exists if the spectral radius of $T_{I_iI_i}$ is less than 1.

Using Eqs. (\ref{eq:ITII}) and (\ref{eq:ITII2}), the recursions can be explicitly solved as
\begin{eqnarray}
\label{eq:brasol}
\bra{O_{i+1}} &=& \bra{O_0}\prod_{k=0}^{i}U_{k,k+1},\\
\ket{O_i} &=& \prod_{k=i}^{N}U_{k,k+1}\ket{O_{N+1}}\label{eq:ketsol}
\end{eqnarray}
where 
\begin{equation}
  \label{eq:umat}
  U_{i,i+1} =
T_{O_{i} O_{i+1}} + T_{O_{i} I_{i}}(\mathbf{I} - T_{I_{i} I_{i}})^{-1}T_{I_{i} O_{i+1}}.
\end{equation}
Finally, the total partition function is obtained as
\begin{equation}
\Xi = \braket{O_0|\prod_{k=0}^{N}U_{k,k+1}|O_{N+1}}.\label{eq:ztot}
\end{equation}

\subsection{Expected densities}
\label{sec:expdensity}
Let us now compute the expected densities.
From the definition of the partition function (Eq. \ref{eq:partition-function}),
the following equalities hold for single-site and bonded pair densities:
\begin{eqnarray}
T\frac{\partial \ln \Xi}{\partial \mu_{S}(a)}  &=& n_{S}(a),\\
T\frac{\partial \ln \Xi}{\partial J_{SS'}(a,b)}   & = & n_{SS'}^b(a,b).
\end{eqnarray}
By explicitly calculating the left-hand sides of these equations using Eq. (\ref{eq:ztot}), we have, for $S = O_i, I_i$ and $S' = O_{i+1}, I_i$,
\begin{eqnarray}
  \label{eq:marginal1c}
  {n}_{S}(a) &=& \frac{\braket{S|a}\braket{a|S}}{\Xi},\\
  \label{eq:marginal2c}
  {n}_{SS'}^b(a,b) &=& \frac{\braket{S|a}\braket{a|T_{SS'}|b}\braket{b|S'}}{\Xi}.
\end{eqnarray}
It is readily proved that these expressions satisfy the relations between bonded pair and single-site densities (Eqs. \ref{eq:relb1} -- \ref{eq:relb2}).

It is also possible to derive an analytical expression for the expected non-bonded pair densities from
\begin{equation}
T^2\frac{\partial^2\ln \Xi}{\partial\mu_S(a)\partial\mu_{S'}(b)} = n_{SS'}^{nb}(a,b) - n_{S}(a)n_{S'}(b).\label{eq:anacov}
\end{equation}
That is,
\begin{equation}
  \label{eq:marginalnb}
  n_{SS'}^{nb}(a,b) = \frac{\braket{S|a}\braket{a|\Xi_{SS'}|b}\braket{b|S'}}{\Xi}
\end{equation}
where
\begin{eqnarray}
  \Xi_{O_iO_j} &=& \prod_{k=i}^{j-1}U_{k,k+1},\\
  \Xi_{O_iI_j} &=& \Xi_{O_iO_j}T_{O_jI_j}(\mathbf{I} - T_{I_jI_j})^{-1},\\
  \Xi_{I_iO_j} &=& (\mathbf{I} - T_{I_iI_i})^{-1}T_{I_iO_{i+1}}\Xi_{O_{i+1}O_j},\\
  \Xi_{I_iI_j} &=& (\mathbf{I} - T_{I_iI_i})^{-1}T_{I_iO_{i+1}}\Xi_{O_{i+1}I_j}.
\end{eqnarray}
However, Eq. (\ref{eq:marginalnb}) is not used in practice for the reason described below (Section \ref{sec:kmat}). This expression should be considered as an artifact of the present approximation on the one-dimensional lattice system. In fact, under the mean-field approximation, one should have $n_{SS'}^{nb}(a,b) = n_{S}(a)n_{S'}(b)$, but this does not hold for Eq. (\ref{eq:marginalnb}).
\subsection{Thermodynamic functions}
Several ``thermodynamic functions'' are defined for quantifying the stability of the system under perturbations. First, the free energy function
\begin{equation}
  \Omega = -T\ln \Xi\label{eq:free-energy}
\end{equation}
should be regarded as a grand potential because alignments of varying lengths are considered in the ensemble. This free energy is a measure of the likelihood of
 alignments expressed in terms of the number densities.
By rearranging Eq. (\ref{eq:boltzmann}) and averaging over
all alignments, the free energy can be decomposed as
\begin{equation}
  \label{eq:F-decomp}
  \Omega = U - TS - G
\end{equation}
where $U$, $S$ and $G$ are the internal energy, entropy and Gibbs energy of the system.
The internal energy of the system is given as
\begin{equation}
  \label{eq:internal-energy}
  U = U_{b} + U_{nb}
\end{equation}
where $U_{b}$ and $U_{nb}$ are bonded and non-bonded energies, respectively, defined (under the mean-field approximation) by
\begin{eqnarray}
  U_b &=&  -{\sum_{(S,S')}^{\mathrm{b.p.}}}\sum_{a,b}J_{SS'}(a,b)n_{SS'}^b(a,b),\\
  U_{nb} &=& -\frac{1}{2}\sum_{S}\sum_{a}\tilde{K}_{S}(a)[n_{S}(a) - \bar{n}_S(a)].
\end{eqnarray}
These correspond to the first two terms on the right-hand side of Eq. (\ref{eq:energy}).
The internal energy represents the mean ``direct'' interactions (bonded and non-bonded) between sites.
The Gibbs energy is defined as
\begin{equation}
  \label{eq:gibbs-energy}
  G = \sum_{S,a}\mu_{S}(a)n_{S}(a),
\end{equation}
and this quantity represents the work exerted by the chemical potential to maintain the single-site densities.
Finally, the entropy is given as
\begin{equation}
  \label{eq:entropy2}
  S = (\Omega - U + G)/T
\end{equation}
which is equivalent to the entropy in Eq. (\ref{eq:entropy}) and thus is a measure of randomness of the alignments.

The temperature $T$ is set to 1 and the chemical potentials are set to 0 for all $S \in \mathcal{S}, a \in \mathcal{A}_S$ when the parameters $J$ (and $K$) are determined. This state is referred to as the reference state in the following.
\subsection{Gauge fixing}
\label{sec:gauge}
The relations among the densities (Eqs. \ref{eq:rel1}--\ref{eq:relnb2}) indicate that not all the parameters, $\mu$, $J$, and $K$, are independent. When determining or changing the model parameters, we may therefore fix some of them to arbitrary values without losing generality. From the normalization condition (Eq. \ref{eq:rel1}) of core sites, it is always possible to set 
\begin{equation}
\mu_{O_{i}}(\mathtt{-}) = 0
\end{equation}
for all the sites $O_i \in \mathcal{O}$ (``$\mathtt{-}$'' stands for the deletion). From this and the relations Eqs. (\ref{eq:relb1}) and (\ref{eq:relb2}),
it is always possible to set
\begin{equation}
J_{O_{i}O_{i+1}}(\mathtt{-},\mathtt{-}) = 0.
\end{equation}
Although there are other degrees of freedom that can be also fixed, they are not relevant to the present study so I will not fix them.

Furthermore, at the reference state, I set all $\mu_S(a)$ to zero. This is possible because any values of $\mu_{S'}(b)$ may be absorbed into $J_{SS'}(a,b)$ when
determining the parameters (c.f., Eq. \ref{eq:tmat}).
Following the convention of Morcos et al.~\cite{MorcosETAL2011}, I also set
\begin{eqnarray}
K_{O_iO_{j}}(\mathtt{-},b) & = & K_{O_{i}O_{j}}(a,\mathtt{-}) = 0, \label{eq:KOzero}
\end{eqnarray}
for all $a \in \mathcal{A}_{O_i}$ and $b \in \mathcal{A}_{O_j}$.

\section{Materials and Methods}
\label{sec:Materials}
\subsection{Data preparation and determining lattice structure}
I have downloaded the MSA's and profile HMM's for the globin (PF00042) and (immunoglobulin) V-set (PF07686) domains from the Pfam database (version 28)~\cite{Pfam}. For the globin domain, the full alignment of 17,947 amino acid sequences were used. For the V-set domain, the full alignment of of 23,976 sequences was used. In addition, I have downloaded 17 families from the top 20 largest Pfam families with the model length of less than 300 sites. For these 17 families, the representative set  of alignments (with 75\% sequence identity cutoff) were used due to the large size of the alignments.

In the present study, the lattice structure of a MSA was derived from the corresponding Pfam model. That is, each core site corresponds to a profile HMM match state, and each insert site to a profile HMM insert state. 

\subsection{Observed densities}
The simplest way to estimate the single-site, bonded and non-bonded pair densities from a MSA of $M$ sequences is to approximate $P(\mathbf{X}) = 1/M$ for all the $M$ sequences. In practice, I used pseudo-counts as well as sequence weights as in Morcos et al.\cite{MorcosETAL2011} to improve the robustness of the estimates.
Let there be $M$ aligned sequences, $\{\mathbf{X}^t\}_{t=1,\cdots,M}$, in a given MSA and suppose the structure of the lattice system has been set. The observed densities are defined as follows:
\begin{eqnarray}
\bar{n}_{S}(a) &=& C\left[\frac{\gamma}{q_S}+\sum_{t=1}^M\frac{n_{S}(a|\mathbf{X}^t)}{m_t}\right],\\
\bar{n}_{SS'}^s(a,b) &=& C\left[\frac{\gamma}{2q_Sq_{S'}}+\sum_{t=1}^M\frac{n_{SS'}^s(a,b|\mathbf{X}^t)}{m_t}\right],\\
\bar{n}_{SS'}^l(a,b) &=& C\left[\frac{\gamma}{q_Sq_{S'}}+\sum_{t=1}^M\frac{n_{SS'}^l(a,b|\mathbf{X}^t)}{m_t}\right]
\end{eqnarray}
where $S \in \mathcal{S}$, $q_S = |\mathcal{A}_S|$, $\gamma$ is the pseudo-count, $m_t$ is the number of sequences in the MSA that are highly homologous ($> 80$\% sequence identity) to the sequence $t$, and $C = 1/(\gamma + \sum_t 1/m_t)$ with $\gamma = 0.1\sum_t1/m_t$.
Note that these estimated densities satisfy the relations analogous to Eqs. (\ref{eq:rel1})--(\ref{eq:relnb2}).

\subsection{Determining the $J$ matrices}
\begin{figure}
   \begin{center}
     \includegraphics[width=0.5\textwidth]{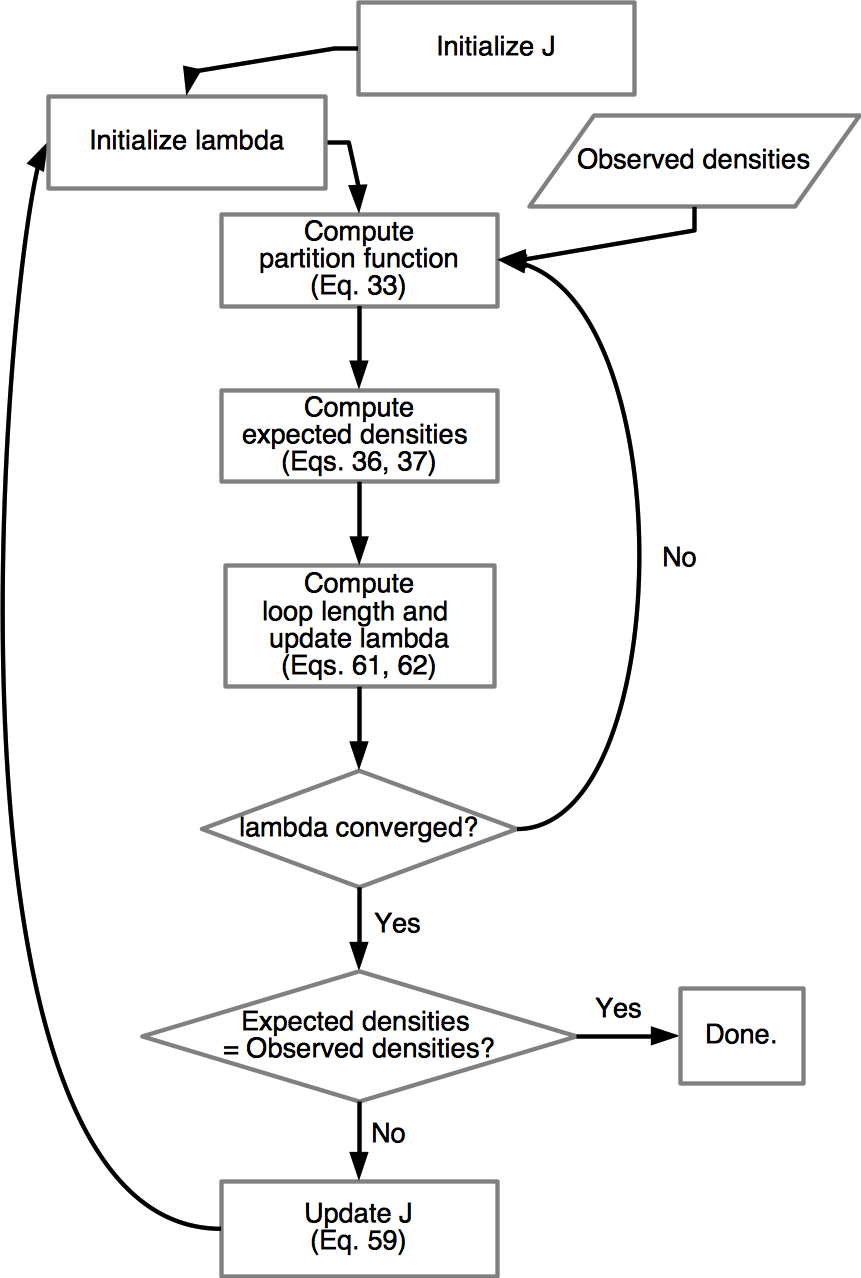}
   \end{center}
   \caption{\label{fig:detJ}
     Flow chart for determining the $J$ matrix parameters.
   }
\end{figure}
As mentioned above, the temperature is set to unity ($T = 1$)
in the process of parameter determination. To determine $J$, Eq. (\ref{eq:marginal2c}) is rearranged to have
\begin{equation}
  \label{eq:detJ}
J_{SS'}(a,b) = \log\left[\frac{{n}_{SS'}^b(a,b)\Xi}{\braket{S|a}\braket{b|S'}}\right]
\end{equation}
where it is assumed $\mu_{S'}(b) = 0$ and $\tilde{K}_{S'}(b) = 0$ for all $S'\in \mathcal{S}$ and $b\in\mathcal{A}_{S'}$ (see Section \ref{sec:gauge}).
Setting $\tilde{K}_{S'}(b) = 0$ is possible because the expected number densities are set to the observed values (see Eq. \ref{eq:mean-field}).
By replacing ${{n}_{SS'}^{b}(a,b)}$ with the observed value ${\bar{n}_{SS'}^b(a,b)}$, one can iteratively update the values of $J$ and compute the partition function until this equation actually holds. In practice, a relaxation parameter is introduced $\alpha$ to improve the stability of convergence.
Thus, from the $\nu$-th step of iteration, the next updated value is
obtained by the following scheme.
\begin{eqnarray}
J_{SS'}'(a,b) &=& \log\left[\frac{\bar{n}_{SS'}^b(a,b)\Xi^{(\nu)}}{\braket{S^{(\nu)}|a}\braket{b|S'^{(\nu)}}}\right],\\
J_{SS'}^{(\nu + 1)}(a,b) &=& (1 - \alpha) J_{SS'}^{(\nu)}(a,b) + \alpha J_{SS'}'(a,b).
\end{eqnarray}
I found the values $\alpha = 0.1 \sim 0.3$ were effective.

Determining $J_{I_iI_i}$ necessitates a special treatment due to the requirement
that the spectral radius of the transfer matrix $T_{I_iI_i}$ must be less than 1 (see Eq. \ref{eq:ITII}). In order to force $\mathbf{I} - T_{I_iI_i}$ to be invertible,
a parameter $\lambda_i > 0$ is introduced
such that $\|T_{I_iI_i}/\lambda_i\| < 1$. Then
Eq. (\ref{eq:marginal2c}) for $S = S' = I_i$ becomes
\begin{equation}
  {n}_{I_iI_i}^b(a,b) = \frac{\braket{I_i|a}\braket{a|T_{I_iI_i}|b}\braket{b|I_i}}{\Xi\lambda_i}.
\end{equation}
Let us define the ``loop length'' $l_i$ as
\begin{equation}
  l_i = \sum_{a,b\in \mathcal{A}_{I_i}}{n}_{I_iI_i}^b(a,b)
\end{equation}
and denote its observed counterpart by $\bar{l}_i$. By imposing $l_i = \bar{l_i}$ we have
\begin{equation}
  \label{eq:lambda}
  \lambda_i = \frac{\bra{I_i}{T_{I_iI_i}}\ket{I_i}}{\Xi\bar{l}_i}
\end{equation}
which is a self-consistent equation for $\lambda_i$. Thus, first $\lambda_i$ is set to a sufficiently large value and compute the partition function and expected densities. Then, $\lambda_i$ is updated by Eq. (\ref{eq:lambda}),
and by using the updated value of $\lambda_i$, we again compute the partition function and expected densities. This process is repeated until the value of $\lambda_i$ converges.
After the convergence of $\lambda_i$ for all $i$, $J_{I_iI_i}$ is updated as in Eq. (\ref{eq:marginal2c}) without including $\lambda_i$. In this way, the contribution of $\lambda_i$ is incorporated into the updated value of $J_{I_iI_i}$, and $\lambda_i$ will eventually converge to 1, and hence may be omitted in later calculations.

The overall procedure for determining the $J$ matrix is shown in Fig. \ref{fig:detJ}. In this procedure, the given data are the observed densities and initial values for $J$ and $\lambda_i$. After the partition function and expected densities are computed, $\lambda_i$ is iteratively updated. After $\lambda_i$ has converged, $J$ is updated. Convergence is checked based on the difference of the expected bonded pair densities from their observed values: when the root mean squeare between the two densities is less than $10^{-13}$, the iteration is stopped.

\subsection{Determining the $K$ matrix}
\label{sec:kmat}
In this study, only those between core sites are taken into account for non-bonded interactions. Including non-bonded interactions with insert sites is numerically unstable because the spectral radius of $T_{I_iI_i}$ may easily exceed 1.
Noting the gauge fixing (Eq. \ref{eq:KOzero}), we first determine $K_{O_iO_j}(a,b)$ viewed as a $20N\times 20N$ matrix (consisting of $N\times N$ blocks of $20\times 20$ submatrices) by discarding the rows and columns including deletion. Then, by fixing the values of $K$, we determine the $J$ matrices.

Let the observed covariance matrix of single-site counts be $C$:
\begin{equation}
  C_{O_iO_j}(a,b) = \bar{n}_{O_iO_j}^{nb}(a,b) - \bar{n}_{O_i}(a)\bar{n}_{O_j}(b).
\end{equation}
In a similar manner as in Morcos et al.\cite{MorcosETAL2011}, one could apply the Plefka expansion~\cite{Plefka1982,Kappen98boltzmannmachine,MorcosETAL2011} to the grand potential (Eq. \ref{eq:free-energy}) with $K=0$ as the reference state.
However, I found that thus obtained $K$ made the system unstable under
very weak perturbations. This behavior is perhaps due to the incompatibility of the mean-field approximation with the one-dimensional system (see the remark at the end of Section \ref{sec:expdensity}).
In order to cope with this problem, I employ the following Gaussian (harmonic) approximation.
By assuming the single-site densities are Gaussian random variables yielding
the observed covariance, the non-bonded coupling is given as
\begin{equation}
K = - C^{-1}, \label{eq:K}
\end{equation}
which is identical to that derived by Morcos et al.~\cite{MorcosETAL2011}
using the Plefka expansion, except for the diagonal blocks (i.e., $K_{O_iO_i}$). Unlike their case (where the diagonal blocks are defined to be zero), I use the expression for $K$ as in Eq. (\ref{eq:K}) including the diagonal blocks.
The system was again found to be unstable when the diagonal blocks
(and those for bonded pairs) of $K$ were set to zero.
This approximation makes the $K$ matrix negative semi-definite so that the
observed single-site densities are the most stable ones and there are no other optima as far as non-bonded pairs are concerned.

\subsection{Self-consistent solutions with fixed parameters}
To obtain a self-consistent solution for the recursion equation (Eqs. \ref{eq:forwardO}--\ref{eq:backwardI}) with a given set of parameters $\mu$, $J$ and $K$,
we first set the mean-field $\tilde{K}_S(a) = 0$ for all $S$ and $a$. Then compute the partition function and the expected densities $n_S(a)$ and update
$\tilde{K}_S(a)$ by Eq. (\ref{eq:mean-field}). This process is repeated until convergence. In practice, however, I do not use this self-consistent solution (see below).

\subsection{Self-consistent solutions with fixed sequence length}
Note that our partition function is that of a grand canonical ensemble so the total number of particles (residues) can vary. In practice, however, it is preferable to fix the sequence length for comparing different conditions to be meaningful. This can be achieved by adjusting the chemical potentials. First, let us define the sequence length as the number of particles in the system:
\begin{equation}
L = \sum_{S\in\mathcal{S}}\sum_{a=1}^{20}n_{S}(a).\label{eq:slen}
\end{equation}
Note that the deletion is not included here (i.e., $a = 21$ when $S = O_i$).
Let $\bar{L}$ denote the target sequence length (a constant). At every step of self-consistent calculation, update the chemical potential of each residue (except for deletion) by
\begin{equation}
  \label{eq:lenupdate}
  \mu_{S}(a) = \mu_{S}(a) + \epsilon(\bar{L} - L)
\end{equation}
where $\epsilon$ is a small positive constant ($\epsilon \approx 0.001$).
\begin{figure}
   \begin{center}
     \includegraphics[width=0.5\textwidth]{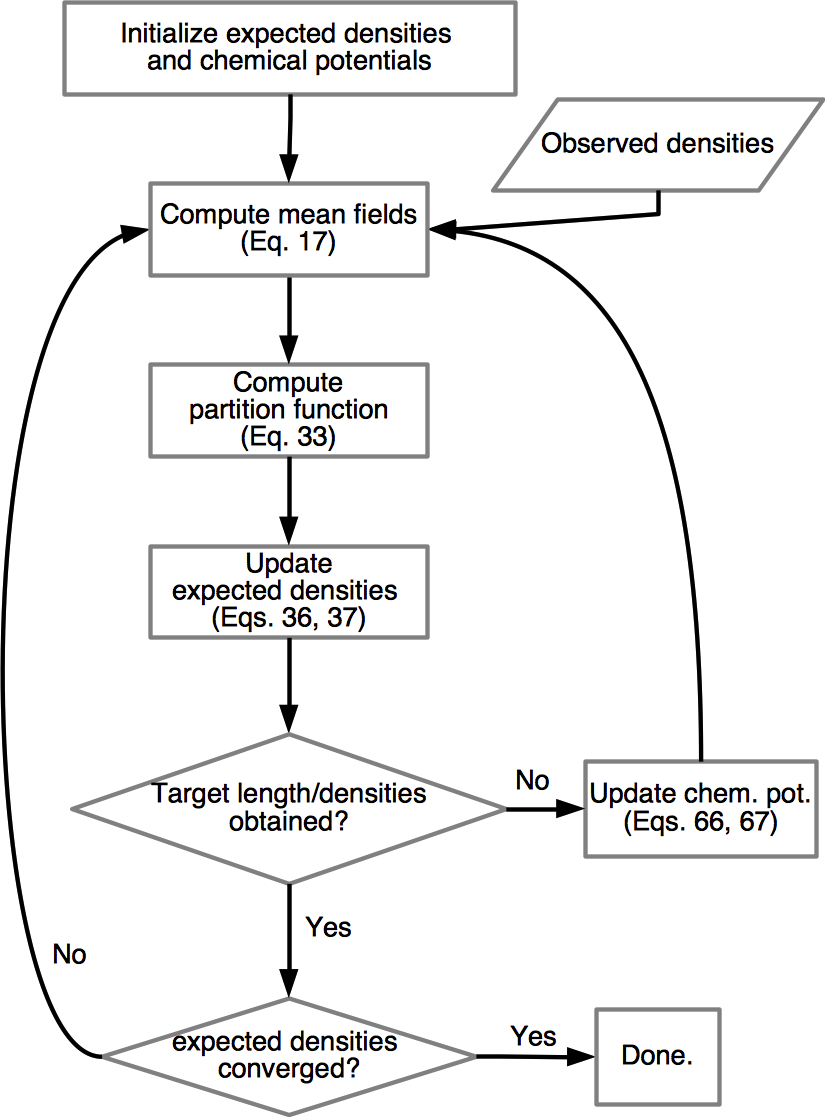}
   \end{center}
   \caption{\label{fig:SCF}
     Flow chart for obtaining the self-consistent solution with fixed sequence length (and fixed single-site densities).
   }
\end{figure}

\subsection{Self-consistent solutions with fixed single-site densities}
In virtual alanine scanning experiments,  the single-site densities of particular sites is specified. Given densities $\hat{n}_S(a)$ for all $a\in\mathcal{A}_S$ for a particular site $S$ can be specified by adjusting the chemical
potentials at every iteration of the self-consistent calculations:
\begin{equation}
  \label{eq:mu-update}
  \mu_{S}(a) = \mu_{S}(a) + \epsilon'[\hat{n}_S(a) - n_{S}(a)]
\end{equation}
where $\epsilon'$ is a positive constant ($\epsilon' \approx 10$). For the case of core sites, it is always possible to set $\mu_S(\mathtt{-}) = 0$ by subtracting this value from those of other residue types of the same site. When the sequence length is to be fixed as well, both Eqs. (\ref{eq:lenupdate}) and (\ref{eq:mu-update}) are applied.

\subsection{Measures of site conservation and difference}
A measure of site conservation is the site entropy~\cite{MacKayInformationTheory} defined by
\begin{equation}
  \label{eq:core-entropy}
  H_{O_i} = -\sum_a \bar{n}_{O_i}(a)\ln{\bar{n}_{O_i}(a)}
\end{equation}
for the reference state. The more well-conserved a site, the lower the value of the site entropy.
The difference between the reference state and a perturbed state is measured by 
the Kullback-Leibler divergence~\cite{MacKayInformationTheory}:
\begin{equation}
  \label{eq:core-div}
  D_{O_i} = \sum_a n_{O_i}(a)\ln\frac{n_{O_i}(a)}{\bar{n}_{O_i}(a)},
\end{equation}
and the total divergence is defined by
\begin{equation}
  \label{eq:tot-core-div}
  D = \sum_{i=1}^{N}D_{O_i}.
\end{equation}

\section{Results}
\label{sec:Results}
I now study the behavior of the lattice gas model of multiple sequence alignment by varying temperature or by ``mutating'' a site. I mostly focus on the effect of non-bonded interactions in the following. For this purpose, I compare the system including both the bonded and non-bonded interactions (referred to as the ``$J+K$'' system in the following) with that including only the bonded interactions (the ``$J$-only'' system). The calculations for the $J$-only system were performed by simply discarding the mean-field, which is justified due to the present definition of the mean-field (Eq. \ref{eq:mean-field}).

All the calculations in the following are based on the ``fixed-length'' solution, and the sequence length (Eq. \ref{eq:slen}) was constrained to that of the reference state.

\subsection{Temperature scanning}
Note that the present model does not exhibit phase transition due to the Gaussian approximation of the non-bonded pair interactions. That is, the $K$ matrix is negative semi-definite so that there exists one and only one minimum for the non-bonded interactions (i.e., at the observed single-site densities).
Nevertheless, solving the self-consistent equation with varying temperatures
 helps to understand the behaviors of interactions. At high temperatures, all the interactions are effectively weakened. This can be regarded as an idealization of uniform random mutations along the protein sequences of the given family. By observing the residue compositions perturbed by increased temperature, we can see which sites are more robust under the perturbations.

 \subsubsection{Globin domain}
 
 \begin{figure*}
   \begin{center}
     \includegraphics[width=\textwidth]{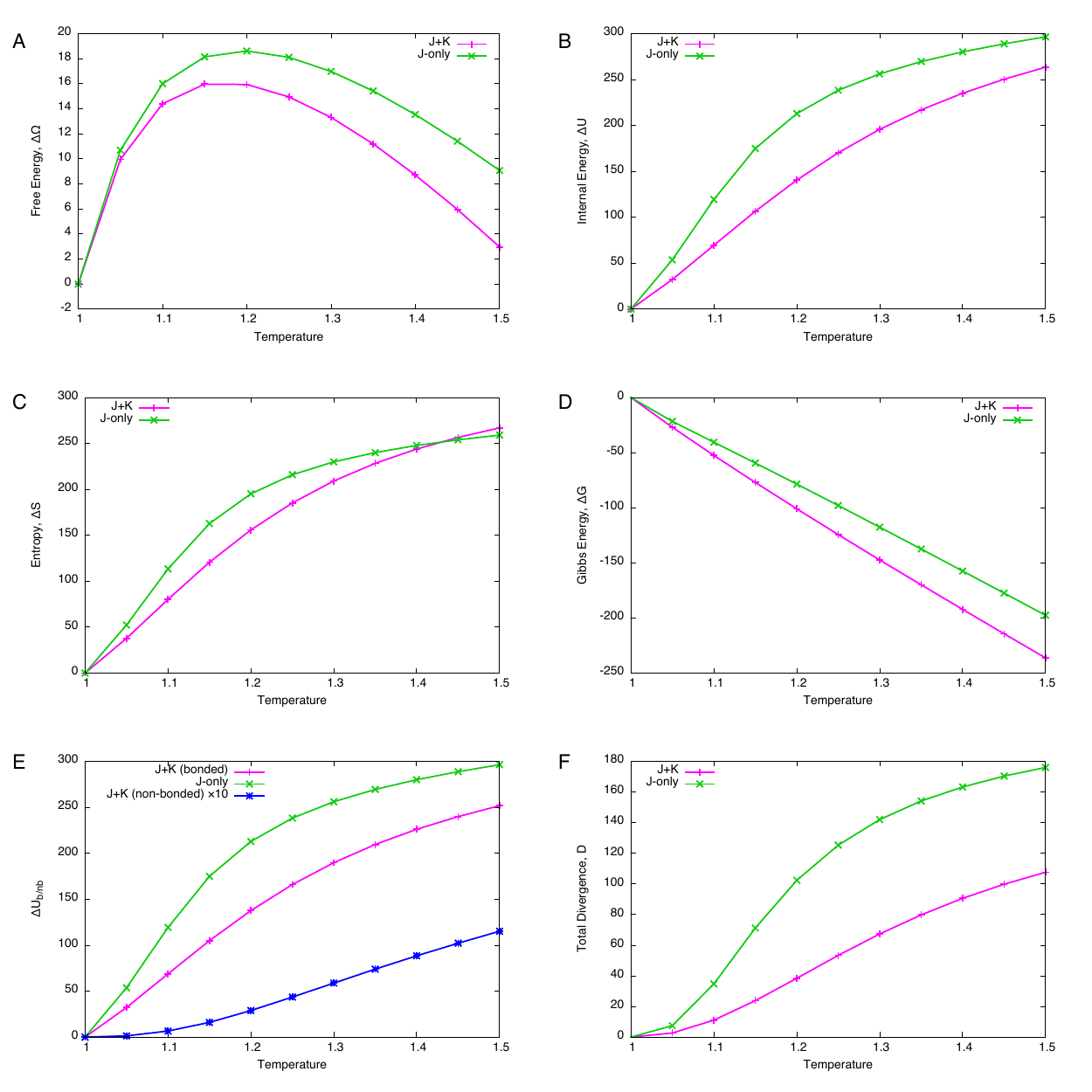}
   \end{center}
   \caption{\label{fig:t-globin}
     Temperature scanning of the globin domain.
     (A) Free energy difference $\Delta \Omega$ from the reference state ($T=1$).
     (B) Internal energy difference $\Delta U$.
     (C) Entropy difference $\Delta S$.
     (D) Gibbs energy difference $\Delta G$.
     (E) Decomposition of internal energy difference into bonded and non-bonded energy differences. The value of non-bonded energy difference (blue line) is multiplied by 10.
     (F) Total divergence of the core site compositions from the reference state (c.f., Eq. \ref{eq:tot-core-div}).
   }
 \end{figure*}

The globin domains are found in a wide variety of organisms ranging from bacteria to higher eukaryotes. Two of the most famous family members are myoglobins and hemoglobins both of which bind the heme prosthetic group. Structurally, globins belong to the class of all-$\alpha$ proteins, 
The lattice gas model of the globin domain consisted of 110 core sites (excluding the termini) and 111 insert sites.

The self-consistent equation was solved for temperature ranging from $T = 1.0$ to $T = 1.7$. Above the latter temperature, the solution could not be obtained stably because the spectral radius of some $T_{I_iI_i}$ exceeded 1.

As the temperature increases, the free energy (grand potential, Eq. \ref{eq:free-energy}) increases up to
around $T = 1.15$ and then it starts to decrease (Figure \ref{fig:t-globin}A). Decomposing the free energy (Eq. \ref{eq:F-decomp}) shows that both the internal energy (Figure \ref{fig:t-globin}B) and entropy (Figure \ref{fig:t-globin}C) increase with temperature. On the other hand, the Gibbs energy (Eq. \ref{eq:gibbs-energy}) monotonically decreases with increasing temperature (Figure \ref{fig:t-globin}D), indicating that the sequence length tends to be longer for higher temperature.
This can be understood from the definition of the transfer matrix $T_{I_iI_i}$.
Since $\|T_{I_iI_i}\| < 1$ is required, $J_{I_iI_i}(a,b) < 0$ holds for all $a,b \in \mathcal{A}_{I_i}$ ($I_i\in \mathcal{I}$) so the increased temperature
potentially allows a larger number of residues to reside at insert sites. In order to fix the sequence length, the chemical potential must be negative, and hence the negative Gibbs energy. 

The behaviors of the $J+K$ and $J$-only systems appear similar regarding the free energy, internal energy, entropy and Gibbs energy. To see the effect of non-bonded interactions more closely, the internal energy was decomposed into bonded interactions and non-bonded interactions for the $J+K$ system (Figure \ref{fig:t-globin}E). It appears that the increase in non-bonded energy is more than an order of magnitude smaller (Figure \ref{fig:t-globin}E, blue line) compared to that of bonded energy (Figure \ref{fig:t-globin}E, magenta line).
Furthermore, the divergence (difference of residue distributions from the reference state) shows a relatively large difference between the $J+K$ and $J$-only systems (Figure \ref{fig:t-globin}F). Thus, the non-bonded interactions are very stable under increased temperatures, and they greatly stabilize the residue composition.

\begin{figure*}
\begin{center}
\includegraphics[width=\textwidth]{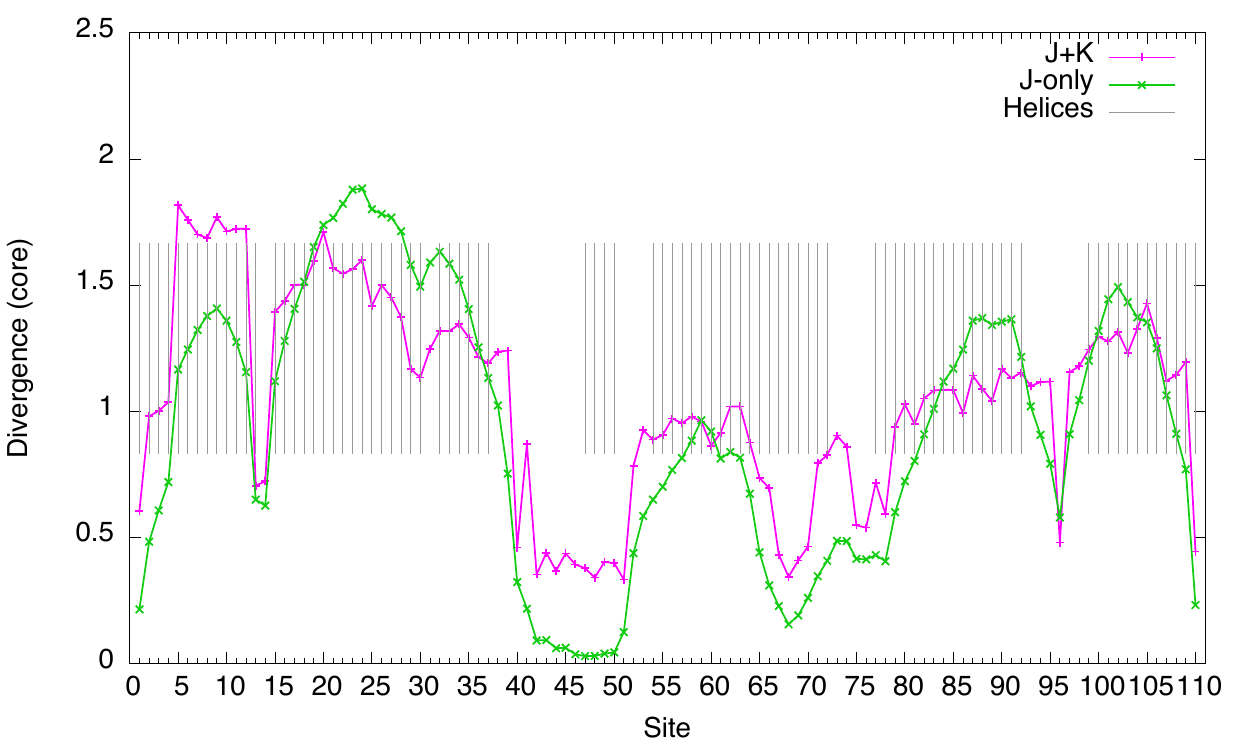}
\end{center}
\caption{\label{fig:t-globin-div}
  Divergence of core sites of the globin domain at $T=1.2$ (c.f., Eq. \ref{eq:core-div}). Gray bars indicate sites annotated as
  helices ($\alpha$-helix, ``H'' or $3_{10}$-helix, ``G'') according to the Pfam model annotation (PF00042). The values for the $J+K$ system are multiplied by 3.
}
\end{figure*}

A closer examination of each site (at $T=1.2$) shows that the magnitude of the divergence of the $J$-only system is about three times as large as that of the $J+K$ system (Figure \ref{fig:t-globin-div}). The broad peaks of the divergence roughly correspond to regions of $\alpha$-helices. Furthermore, with non-bonded interactions, finer peaks match the periodicity of the helices (3 to 4 residues) whereas such periodicity is not observed with the $J$-only system.
Thus, non-bonded interactions seem not only to stabilize the residue composition, but to make the composition more specific to the structure of the domain.

\subsubsection{V-set domain}
\begin{figure*}
\begin{center}
\includegraphics[width=\textwidth]{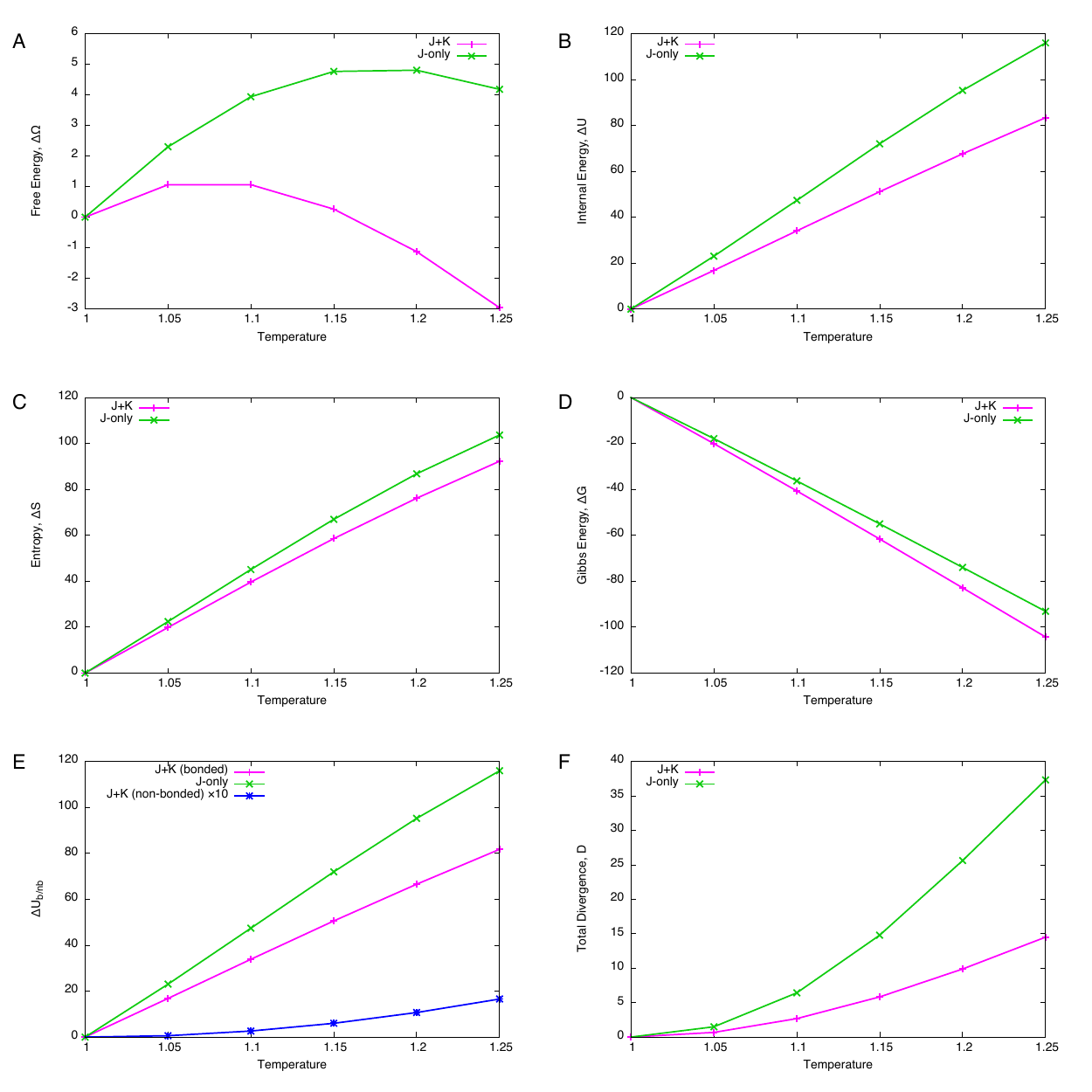}
\end{center}
\caption{\label{fig:t-vset}%
    Temperature scanning of the V-set domain.
  (A) Free energy difference $\Delta \Omega$ from the reference state ($T=1$).
  (B) Internal energy difference $\Delta U$.
  (C) Entropy difference $\Delta S$.
  (D) Gibbs energy difference $\Delta G$.
  (E) Decomposition of internal energy difference into bonded and non-bonded energy differences. The value of non-bonded energy difference (blue line) is multiplied by 10.
  (F) Total divergence of the core site compositions from the reference state (c.f., Eq. \ref{eq:tot-core-div}).
}
\end{figure*}

The V-set domains are found in many proteins the representative members of which are immunoglobulin variable domains. The lattice gas model of this domain consists of 114 core sites (excluding the termini) and 115 insert sites. Structurally, they belong to the all-$\beta$ class having a $\beta$-sandwich structure.

The same procedures were applied to the V-set domain as the globin domain. In this case, however, self-consistent solutions could be obtained only for temperatures $T \leq 1.25$. This may be due to a long insertion allowed at the insert site $I_9$ (average length of 23.5 residues). 
Other than this limitation, the results were found to be qualitatively similar to the case of globins (Figure \ref{fig:t-vset}A-D). However, the free energy decrease is more pronounced for the $J+K$ system, compared to the case of the globin.
Again, while the increase in temperature hardly changes the non-bonded energy (Figure \ref{fig:t-vset}E), the difference of the total divergence between the $J+K$ and $J$-only systems is significant.

A close examination of individual sites at $T=1.2$ also indicates that inclusion of non-bonded interactions greatly suppresses the divergence, and broad peaks roughly correspond to secondary structure elements (in this case, $\beta$-strands). With the non-bonded interactions, finer peaks appear to match with the periodicity of $\beta$-strands (2 residues). Therefore, the conclusion drawn for the globin domain applies also to the V-set domain. That is, the non-bonded interactions act to stabilize the residue composition as well as to make composition more specific to the structure of the domain.

\begin{figure*}
\begin{center}
\includegraphics[width=\textwidth]{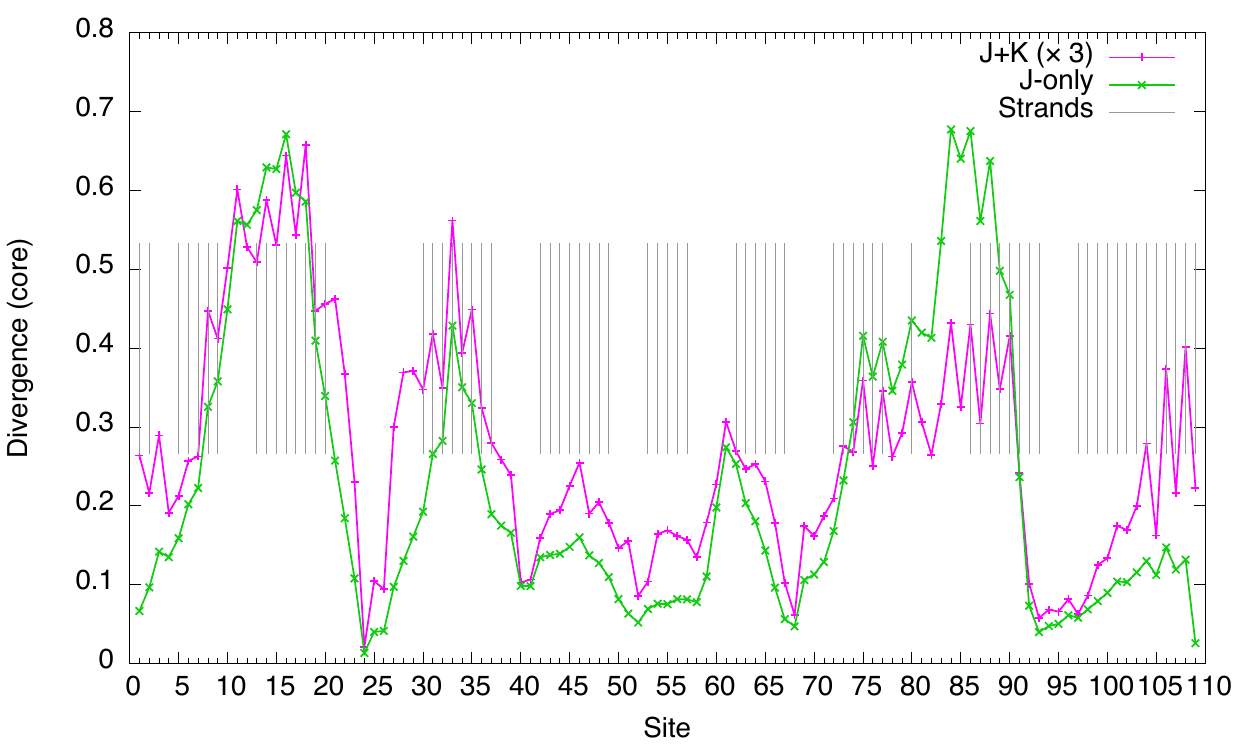}
\end{center}
\caption{\label{fig:t-vset-div}
  Divergence of core sites of the V-set domain at $T=1.2$ (c.f., Eq. \ref{eq:core-div}). Gray bars indicate sites annotated as extended strand, ``E'' according to the Pfam model annotation (PF07686). The values for the $J+K$ system are multiplied by 3.
}
\end{figure*}

\subsection{Alanine scanning}
As opposed to global perturbations such as increased temperature, local perturbations helps us to examine the contribution of individual sites. Local perturbations can be imposed by biasing the residue composition at a site of interest. In this subsection,  the composition of a particular core site was biased in such a way that single-site density was set to 0.95 for alanine and to 0.0025 for all other residue types (including the ``deletion'' residue type). This residue composition can be achieved by adjusting the chemical potential $\mu_{O_i}(a)$.
When the site $O_i$ is constrained in this way, the corresponding equilibrium state is referred to as the ``A$_i$ mutant'' in the following. 

\subsubsection{Globin domain}
\begin{figure*}
\begin{center}
\includegraphics[width=\textwidth]{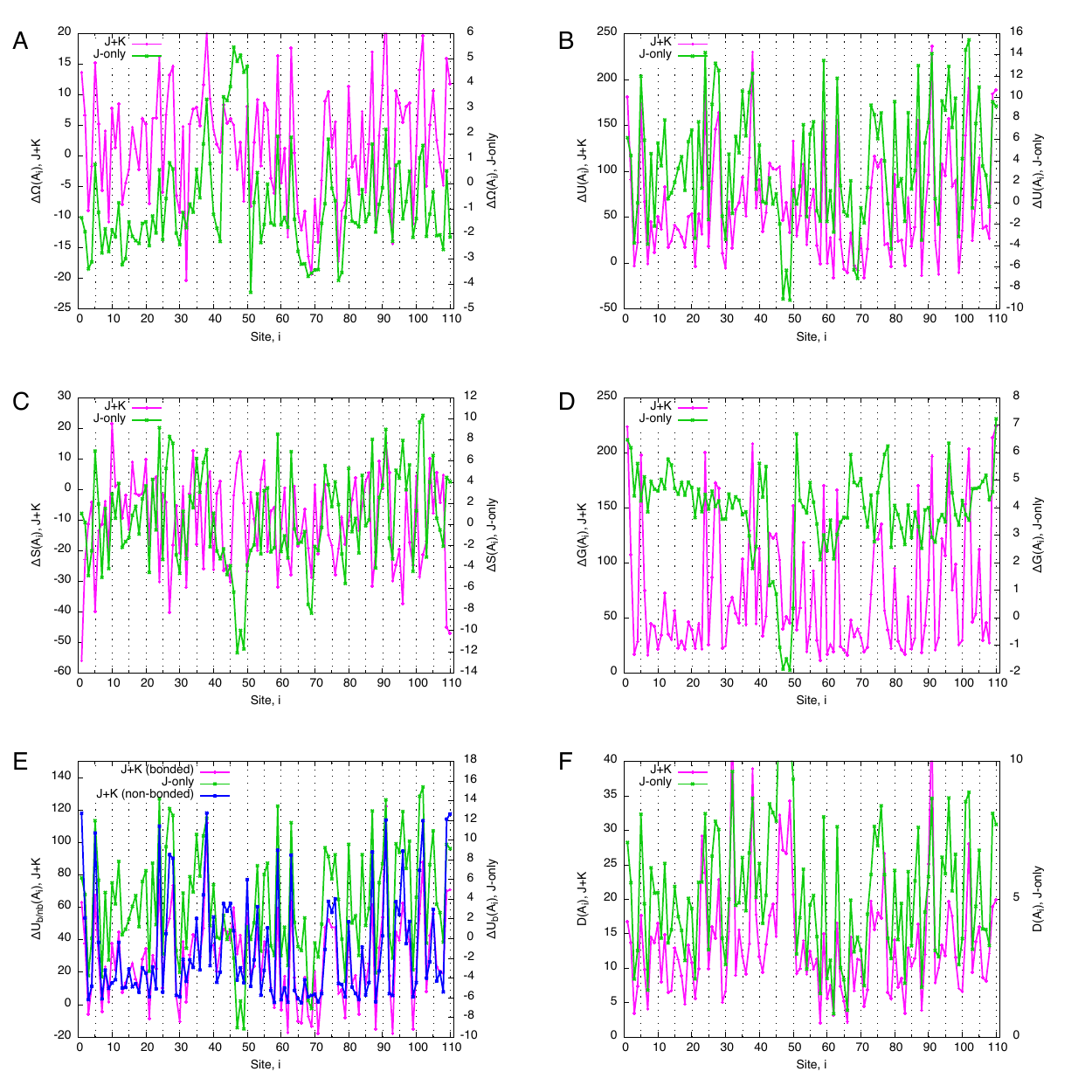}
\end{center}
\caption{\label{fig:a-globin}
  ``Alanine scanning'' of the globin domain. The horizontal axis indicates the site at which the single-site density of a core site was set to 0.95 for alanine, and to 0.0025 for other residue types; the vertical axes indicate associated values (A)-(F), with the $J+K$ system on the left axis, and $J$-only system on the right.
  (A) Free energy difference of ``alanine point mutants'' from the reference state.
  (B) Internal energy difference.
  (C) Entropy difference.
  (D) Gibbs energy difference.
  (E) Decomposed internal energy difference.
  (F) Total divergence of core sites (Eq. \ref{eq:tot-core-div}).
}
\end{figure*}

\begin{figure*}
\begin{center}
\includegraphics[width=\textwidth]{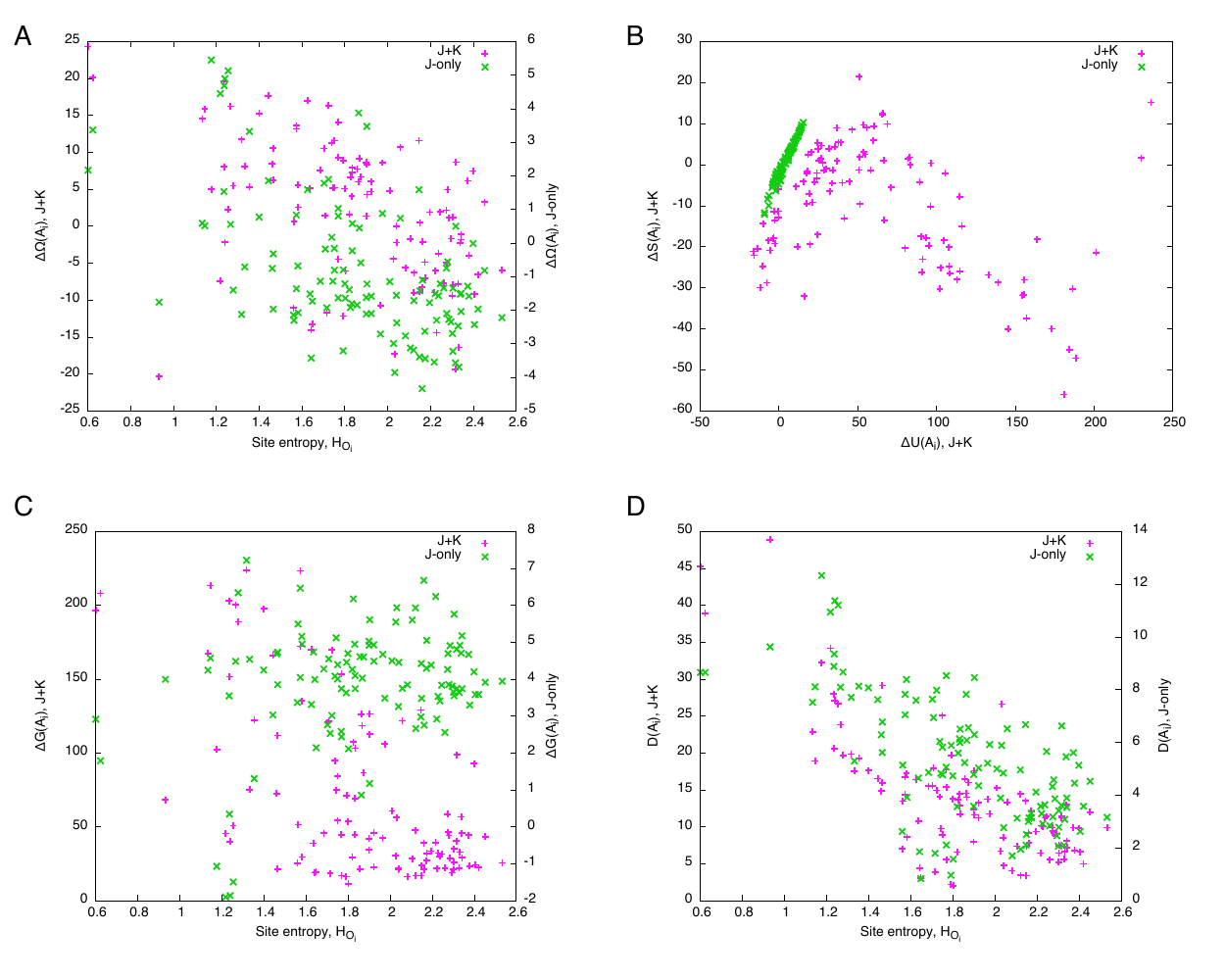}
\end{center}
\caption{\label{fig:a-globin-p}
  Correlations between various quantities for the globin domain.
  (A) Site entropy, $H_{O_i}$, vs. free energy change, $\Delta \Omega$ (left vertical axis for the $J+K$ system, right vertical axis for the $J$-only system).
  (B) Internal energy difference, $\Delta U$, vs. entropy difference, $\Delta S$.
  (C) Site entropy, $H_{O_i}$, vs. Gibbs energy change, $\Delta G$ (left vertical axis for the $J+K$ system, right vertical axis for the $J$-only system).
  (D) Site entropy, $H_{O_i}$, vs. total divergence, $D$ (left vertical axis for the $J+K$ system, right vertical axis for the $J$-only system).
}
\end{figure*}

Comparing the free energy difference between the $J+K$ and $J$-only systems,
it is immediately noticed that the ranges of $\Delta \Omega$ are very different between the two; the former being an order of magnitude larger than the latter.
While a large number of alanine mutants for both the $J+K$ and $J$-only systems (82 and 101, respectively, out of 110) exhibit $\Delta \Omega < 0$ (i.e., favorable mutants), the former ($J+K$) shows a larger number of unfavorable ($\Delta \Omega > 0$) alanine mutants. Apart from the absolute values, the two systems appear to be correlated except for the region from the site 40 to 50 where secondary structures are sparse (c.f., Figure \ref{fig:t-globin-div}). In addition, they seem to be negatively correlated with site entropy (Figure \ref{fig:a-globin-p}A): Highly conserved sites tend to have high $\Delta \Omega$ values (correlation coefficients, CC, were -0.60 and -0.57 for the $J+K$ and $J$-only systems, respectively). Thus, despite the great difference in magnitudes, the $J+K$ system and $J$-only system appear to be similar in terms of free energy difference. Behind this apparent similarity, however, exist different mechanisms, as we shall see in the following.

While internal energy difference, $\Delta U$, also shows a similar correlation
as $\Delta \Omega$ (Figure \ref{fig:a-globin}B), entropy difference exhibits different, somewhat opposite, trends (Figure \ref{fig:a-globin}C). In fact, the relations between the internal energy and entropy are completely different between the $J+K$ and $J$-only systems (Figure \ref{fig:a-globin-p}B). While $\Delta U$ and $\Delta S$ are linearly and positively correlated (CC = 0.99) for the $J$-only system, they relation is more complicated for the $J+K$ system: a positive correlation for $\Delta U < 20$ (CC=0.65) and a negative correlation for $\Delta U > 30$ (CC=-0.69). The region $\Delta U < 20$ corresponds to that spanned by the $J$-only system, and therefore is considered to be the region where local (bonded) interactions are dominant in $\Delta U$. This in turn indicates that a large increase in nonlocal (non-bonded) interactions greatly restricts the residue composition throughout the globin domain. In fact, unlike the case for temperature scanning (Figure \ref{fig:t-globin}), the perturbation by a point mutation induces a large increase in non-bonded energy  that is comparable with that of bonded energy in the $J+K$ system (Figure \ref{fig:a-globin}E).

The Gibbs energy difference, $\Delta G$, reveals a sharp contrast between the two systems (Figures \ref{fig:a-globin}D and \ref{fig:a-globin-p}C). The Gibbs energy differences of the $J+K$ system are clustered below $\Delta G < 50$, but has a long tail towards higher values (skewness was 1.1). On the other hand, those for the $J$-only system are more or less symmetrically distributed around $\Delta G = 5$ (skewness was -1.4). The correlation between $\Delta G$ and site entropy is evident for the $J+K$ system (CC = -0.71), but is nearly absent for the $J$-only system (CC = -0.18) (Figure \ref{fig:a-globin-p}C).

The total divergence shows a trend similar to the Gibbs energy difference in that its values are clustered at lower values and has a long tail towards higher values for the $J+K$ system, and that such is not the case for the $J$-only system (Figure \ref{fig:a-globin}F). Although in the both systems the total divergence is well correlated with site entropy, the correlation is higher for the $J+K$ system (CC = -0.78) than for the $J$-only system (CC= -0.71) (Figure \ref{fig:a-globin-p}D). In the $J$-only system, each mutation perturbs the residue compositions only locally around the mutated site, whereas in the $J+K$ system, a mutation at one site perturbs many sites across the the entire domain. As a result, the contrast between the effects of mutations at highly conserved sites and less conserved sites is higher for the $J+K$ system than for the $J$-only system.

In the globin domain, the two most highly conserved residues are
phenylalanine (Phe) at site 38 ($H_{O_i} = 0.67$) and histidine (His)
at site 91 ($H_{O_i} = 0.64$). The alanine mutants at these sites show
large differences in $\Delta\Omega$ (Fig. \ref{fig:a-globin-p}A), $\Delta U$ (the two points with the largest $\Delta U$ in Fig. \ref{fig:a-globin-p}B) and $\Delta G$ (Fig. \ref{fig:a-globin-p}C).
According to a detailed study by Ota et al.~\cite{OtaETAL1997}, these two residue are conserved for different reasons: Phe at site 38 (``CD1'' in \cite{OtaETAL1997}) is conserved for structural stability whereas His at site 91 (``F8'') is conserved for the heme-binding function at the cost of structural stability.
While it is reasonable to observe that the $A_{38}$ mutant of the structurally conserved Phe significantly disturbs the system, the present result suggests that the $A_{91}$ mutant of the functionally conserved His is also maintained by a significant amount of interactions with other sites. This may indicate the importance of structural scaffold to maintain protein function.

\subsubsection{V-set domain}
\begin{figure*}
\begin{center}
\includegraphics[width=\textwidth]{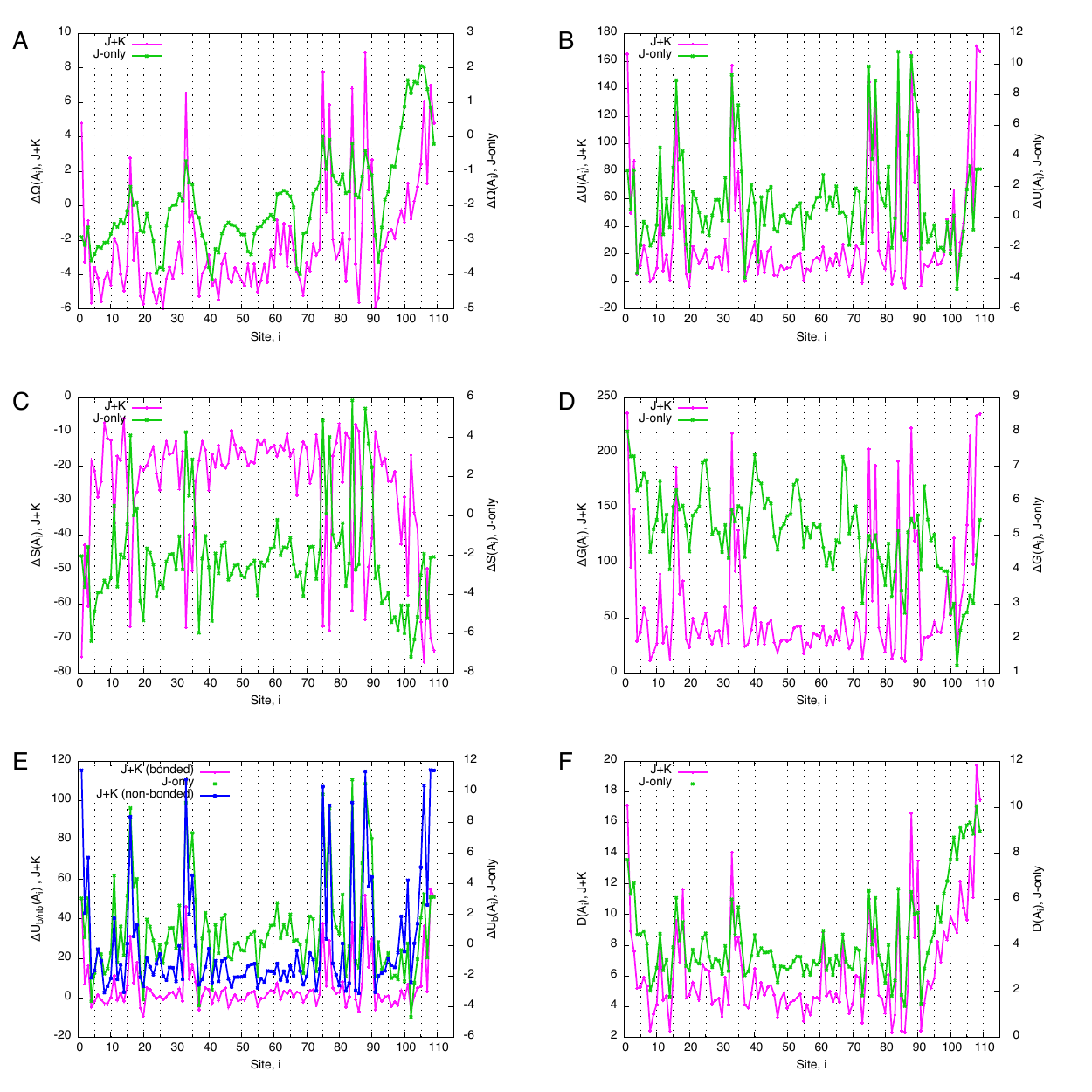}
\end{center}
\caption{\label{fig:a-vset}
  ``Alanine scanning'' of the V-set domain. The horizontal axis indicates the site at which the single-site density of a core site was set to 0.95 for alanine, and to 0.0025 for other residue types; the vertical axes indicate associated values (A)-(F), with the $J+K$ system on the left axis, and $J$-only system on the right.
  (A) Free energy difference of ``alanine point mutants'' from the reference state.
  (B) Internal energy difference.
  (C) Entropy difference.
  (D) Gibbs energy difference.
  (E) Decomposed internal energy difference.
  (F) Total divergence of core sites (Eq. \ref{eq:tot-core-div}).
}
\end{figure*}
\begin{figure*}
\begin{center}
\includegraphics[width=\textwidth]{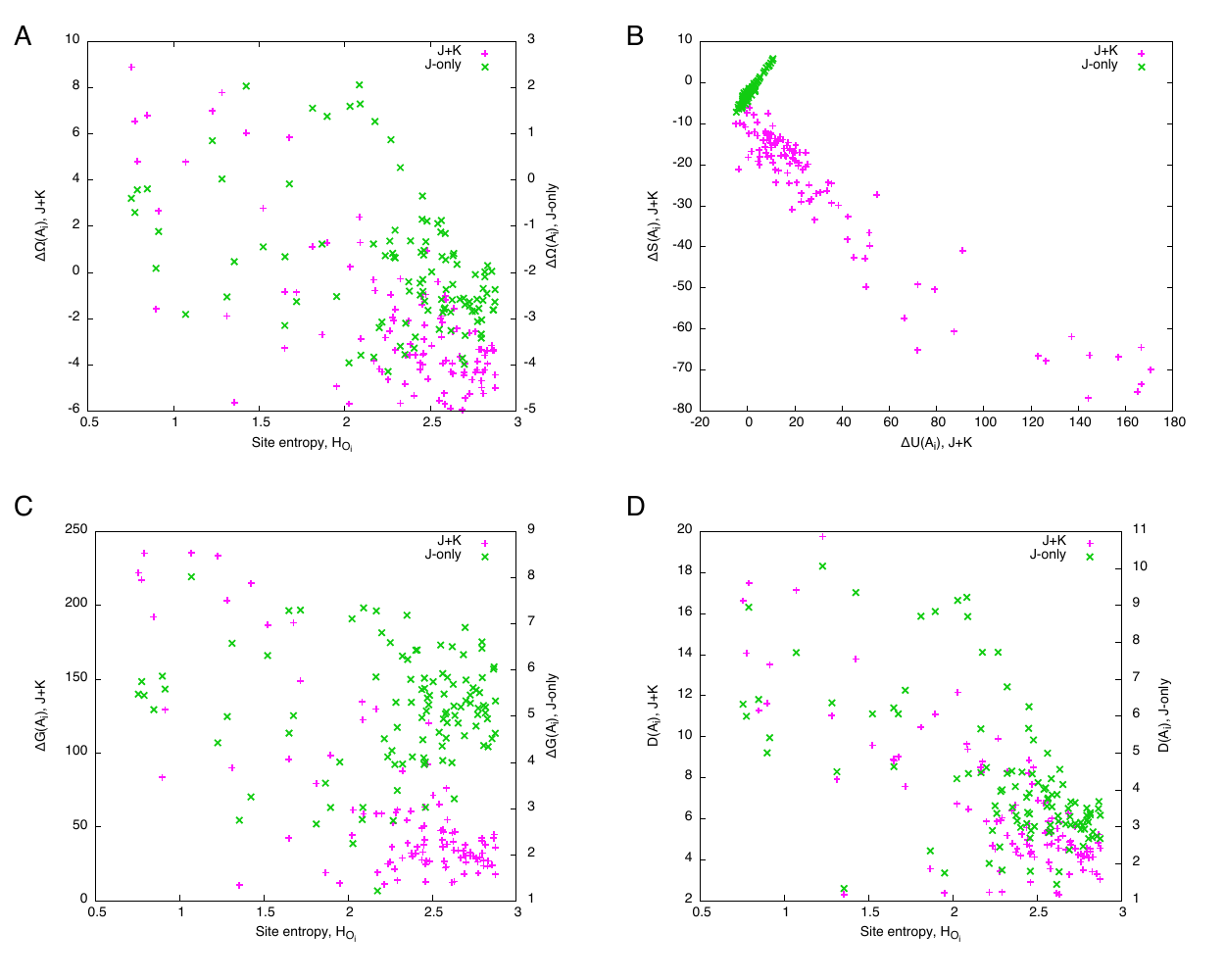}
\end{center}
\caption{\label{fig:a-vset-p}
  Correlations between various quantities for the V-set domain.
  (A) Site entropy, $H_{O_i}$, vs. free energy change, $\Delta \Omega$ (left vertical axis for the $J+K$ system, right vertical axis for the $J$-only system).
  (B) Internal energy difference, $\Delta U$, vs. entropy difference, $\Delta S$.
  (C) Site entropy, $H_{O_i}$, vs. Gibbs energy change, $\Delta G$ (left vertical axis for the $J+K$ system, right vertical axis for the $J$-only system).
  (D) Site entropy, $H_{O_i}$, vs. total divergence, $D$ (left vertical axis for the $J+K$ system, right vertical axis for the $J$-only system).
}
\end{figure*}

The case for the V-set domain is mostly similar to that for the globin domain (Figures \ref{fig:a-vset} and \ref{fig:a-vset-p}).
However, there  are some marked differences to be noted.
First, the free energy differences $\Delta \Omega$ due to alanine mutations
take both positive and negative values for the $J+K$ system, but only negative values for the $J$-only system. The positive values for the former corresponds to relatively well-conserved sites, as can be seen in Figure \ref{fig:a-vset-p}A.
In fact, the correlation between $\Delta \Omega$ and site entropy is significantly higher for the $J+K$ system (CC = -0.72) than for the $J$-only system (CC = -0.52). Second, while the correlation between internal energy and entropy differences
is linear and positive for the $J$-only system (CC = 0.96) as was the case with the globin, that for the $J+K$ system of the V-set domain shows only a negative trend for the entire range of $\Delta U$ (CC = -0.92).
Third, the contrast of the Gibbs energy difference is far more pronounced (Figure \ref{fig:a-vset}D, the skewness was 1.8 for $J+K$ and -0.37 for $J$-only) and its correlation with site entropy is very high for the $J+K$ system (CC = -0.80) whereas it is negligible for the $J$-only system (CC = -0.08) (Figure \ref{fig:a-vset-p}C). Similarly,  as for total divergence, the $J+K$ system shows sharper contrast (Figure \ref{fig:a-vset}F) and higher correlation with site entropy (CC = -0.80, Figure \ref{fig:a-vset-p}D) than the $J$-only system (CC = -0.67).

Thus, compared to the case with the globin, the differences between the $J+K$ and $J$-only systems are more pronounced. This may be due to the difference in the structures of these domains. The globin domain has an all-$\alpha$ fold in which local interactions in $\alpha$-helices are prominent, whereas the V-set domain has an all-$\beta$ fold in which nonlocal interactions between $\beta$-strands are prominent. This difference may be reflected in the non-bonded interactions  of the lattice gas model, hence the pronounced difference between the $J+K$ and $J$-only systems.

\subsubsection{Other protein families}
\begin{figure*}
\begin{center}
\includegraphics[width=\textwidth]{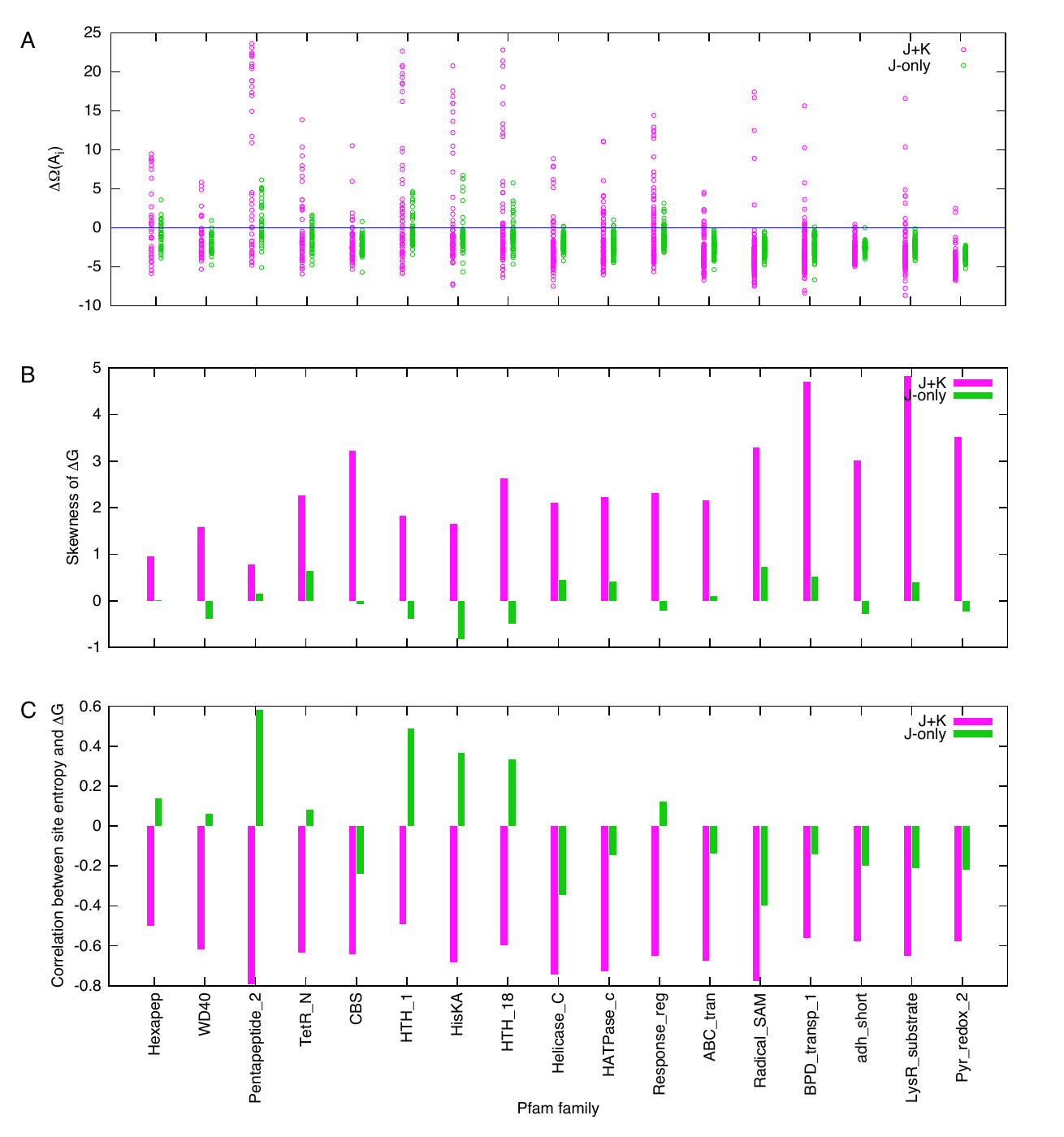}
\end{center}
\caption{\label{fig:top17}
  Alanine scanning of 17 Pfam families.
  (A) Free energy difference (cf. Figs. \ref{fig:a-globin}A and \ref{fig:a-vset}A).
  (B) Skewness (standardized) of $\Delta G$  (cf. Figs. \ref{fig:a-globin}D and \ref{fig:a-vset}D).
  (C) Correlation coefficient between site entropy and Gibbs energy $\Delta G$
  (cf. Figs. \ref{fig:a-globin-p}C and \ref{fig:a-vset-p}C).
}
\end{figure*}
To confirm the observations made above, alanine scanning was performed for 17 Pfam families that are the largest in the number of family members and are of model length of less than 300 sites. The free energy difference, $\Delta\Omega(A_i)$, tends to have more positive values for the $J+K$ system than for the $J$-only system (Fig. \ref{fig:top17A}, cf. Figs. \ref{fig:a-globin}A and \ref{fig:a-vset}C). The skewness (i.e., the standardized third moment) of $\Delta G(A_i)$ consistently have positive values for the $J+K$ system whereas it can be either positive or negative for the $J$-only system (Fig. \ref{fig:top17}B). The negative correlation between site entropy and $\Delta G(A_i)$ was also clear for the $J+K$ system whereas such was not the case for the $J$-only system (Fig. \ref{fig:top17}C). Thus, the trend that the non-bonded interaction enhances correlation with sequence conservation seem to hold generally.

\section{Discussion}
\label{sec:Discussion}

One of the fundamental assumptions of the present lattice gas model is that
alignment sites can be classified into core sites and insert sites.
Although this classification may be ambiguous to some extent, once the
classification is made, the lattice structure is uniquely determined.
While the lattice structure reflects the chemical structure of 
polypeptide chains, interactions between the lattice sites are not limited
to those that are local along the chain. The principle of maximum entropy
allows the model to treat bonded (local) and non-bonded (nonlocal) interactions in a coherent manner. In comparison, the profile HMM~\cite{DurbinETAL} shares a similar lattice structure as the lattice gas model, but it cannot treat nonlocal interactions due to its assumption of the Markov process along the lattice structure. On the other hand, the direct-coupling analysis (as applied to contact prediction)~\cite{MorcosETAL2011}, which casts a MSA as a Potts model~\cite{PottsModel}, simply ignores insert sites so that it cannot faithfully represent polypeptide chains.
Threading methods~\cite{ProteinBioinfo} or conditional random field models~\cite{Kinjo2009} can combine the polypeptide structure with nonlocal interactions, but such integration is often \emph{ad hoc} because there are no well-defined rules or principles for determining the relative contributions of various interactions.
It is possible to treat a MSA without classifying its columns into cores and inserts if one ignores the possibility of adding new sequences in the future. In fact, this approach is adopted by the GREMLIN method by Balakrishnan et al.~\cite{BalakrishnanETAL2011} that is based on the Markov random fields (the present lattice gas model also belongs to this class of statistical models). In practice, however, they discarded columns with excessive  gaps. Such a preprocessing seems to be required because alignments within an insertion are often meaningless. This does not necessarily mean, however, that the existence of the insertion is meaningless. In any case, discarding columns of a MSA will lose the information about the linear chain structure of protein sequences as well as the possibility of adding new sequences without changing the core structure of the MSA.
The present lattice gas model resolves the shortcomings of these previous models
as both bonded and non-bonded interactions as well as insertions naturally emerge from a single framework. The main tricks here are the classification of core and insert sites and the use of residue counts, $n_S(a|\mathbf{X})$ and $n_{SS'}^{b}(a,b|\mathbf{X})$, as fundamental variables rather than the raw alignment sequences ($\mathbf{X}$). These are especially important for treating insert sites where any number of residues are allowed to exist. The lattice gas model can compute the probability of an entire alignment, and what has been conventionally regarded as the probability of residue occurrence at sites should be regarded as the expected number of residues at the sites.

From a theoretical point of view, the present formulation of the lattice gas model offers an interesting perspective regarding the interplay between local and nonlocal interactions. As can be seen from the relations Eqs. (\ref{eq:relb1})--(\ref{eq:relnb2}), or more precisely, from the analogous relations that hold for the number densities, local and nonlocal interactions are not independent of each other, but are related via single-site densities. In this sense, local and nonlocal interactions must be consistent with each other~\cite{Go1983}, and the consistency is inherently embedded in a (well-curated) MSA.
In the conventional formulation of the direct-coupling analysis,
only the relations corresponding to Eqs. (\ref{eq:relnb1}) and (\ref{eq:relnb2}) are present because the chain structure is absent. Since the parameters conjugate to the single-site densities are external fields (chemical potentials in the present case) which are not intrinsic to the system, the relations Eqs. (\ref{eq:relnb1}) and (\ref{eq:relnb2}) alone do not address the consistency between local and nonlocal interactions.

In this study, I have adopted the Gaussian approximation for the non-bonded coupling parameters (Eq. \ref{eq:K}) as well as the mean-field approximation (Eq. \ref{eq:mean-field}) for computing the partition function.
This approach has its advantages and disadvantages. The advantages are that
the parameters are readily obtained and that the
partition function can be computed analytically and efficiently. These enable us to study the system under various perturbations relatively easily.
A major disadvantage is that it is not possible to determine the $K$ matrix self-consistently. I therefore resorted to the Gaussian approximation by implicitly assuming that each site is independent of other sites, which is not fully consistent with the lattice structure of the system. The reason for this inconsistency is likely to be that the assumption for the mean-field approximation (i.e., non-bonded interactions are relatively weak; see references\cite{Plefka1982,MorcosETAL2011}) does not actually hold in the present case.
Due to this approximation, the system does not exhibit a phase transition that might be induced by increased temperatures or by mutations at potentially important sites. 
In addition, the Gaussian approximation required that the diagonal blocks of the $K$ matrix, $K_{O_iO_i}(a,b)$, be used as in Eq. (\ref{eq:K}), otherwise the reference state was found to be unstable.
The diagonal blocks represent self-interactions, and hence, are
purely site-specific quantities. In this sense, they obscure the
mechanism by which the interactions of each site with other sites
induce the residue composition of that site. 
Overcoming these problems would require the direct maximization of the Lagrangian (Eq. \ref{eq:lagrangian}) with respect to the parameters $K_{SS'}(a,b)$ without diagonal (and bonded pair) blocks.
It is also possible to apply other approximate methods such as pseudo-likelihood maximization~\cite{BalakrishnanETAL2011,EkebergETAL2013,KamisettyETAL2013}. 

Despite these limitations in the treatment of non-bonded interactions,
the present results already provided some interesting observations
regarding the role of non-bonded interactions. An increased temperature exerts
a global and unbiased perturbation on the system. In this case, it was found that non-bonded energy did not significantly change compared to the bonded energy (Figures \ref{fig:t-globin}E and \ref{fig:t-vset}E).
This implies that the residue compositions at each site adapt to
the perturbation in a cooperative manner so that they stay stable.
This in turn suggests, at least within the limitation of the approximations,
that a protein family can accommodate a diverse variety of amino acid sequences
as far as the pattern of correlations between sites is conserved.
On the other hand, the virtual alanine scanning revealed a more conspicuous
effect of non-bonded interactions. Alanine mutations at well-conserved sites
disturbed the system to a greater extent as measured by free energy,
Gibbs energy and total divergence (Figures \ref{fig:a-globin-p} and \ref{fig:a-vset-p}), and the relation between internal energy and entropy changes was completely different from those of $J$-only systems. In particular, the observation that many or most of the free energy changes were negative for the $J$-only system (Figures \ref{fig:a-globin}A and \ref{fig:a-vset}A) suggests that residue conservation cannot be explained without considering nonlocal (non-bonded) effects.

The interactions in the lattice gas model originate solely from the statistics of a MSA. They are therefore not directly related to physical interactions. However, it has been demonstrated that the $K$ matrix as used in this study is a good predictor of physical contacts in native protein structures~\cite{deJuanETAL2013,TaylorETAL2013}. To further confirm this,
the present results showed that the effect of non-bonded (statistical) interactions was more pronounced in the V-set domain (an all-$\beta$ fold, involving more nonlocal physical interactions)
than in the globin domain (an all-$\alpha$ fold, involving less nonlocal interactions). In addition, the $J$-only system showed relatively better correlations with conservation for the globin than for the V-set domain, indicating that the bonded interactions also reflect physical local interactions to some extent. This point is also supported by the correlation, albeit weak, between divergence and secondary structures (Figures \ref{fig:t-globin-div} and \ref{fig:t-vset-div}). Thus, the lattice gas model provides a means to connect the information in amino acid sequence with the underlying three-dimensional structure of the domain. This connection cannot be addressed directly in conventional sequence analysis methods such as the profile HMM. In fact, the very existence of long-range correlations indicates that MSA's cannot be modeled as a purely one-dimensional system where long-range correlations simply cannot exist~\cite{Goldenfeld1992}. Considering this fact, it is surprising that conventional multiple sequence alignment methods, inherently based on the one-dimensional system, can produce MSA's
with long-range correlations. This may be a manifestation of the consistency principle indicated above~\cite{Go1983}.

There are a few possible extensions and applications of the present lattice gas model. In the present form, the model is autonomous in the sense that it does not require an input or target sequence for computing various statistical quantities (once the observed statistical quantities are obtained). Nevertheless, it is readily possible to align the model with a particular amino acid sequence to compute a partition function and therefore other quantities conditioned on that input sequence. In this way, the lattice gas model may be used for detecting remote homologs. The present results (e.g., Figures \ref{fig:t-globin-div} and \ref{fig:t-vset-div}) suggest that inclusion of non-bonded interactions would increase the specificity of the alignment. Furthermore, the model can be aligned with a ``sequence'' of a given length with unspecified amino acid residues to compute the partition function that is conditioned on all the amino acid sequences of that length. In this way, one can enumerate those sequences that are compatible with the model. In other words, the model may be used for designing optimal sequences for a given protein family. Such applications may be pursued in the future to open new possibilities in protein sequence analysis.

\begin{acknowledgments}
The author thanks Kentaro Tomii and Sanzo Miyazawa for reading the initial manuscript and providing some references and comments, Haruki Nakamura for the support during the development of this work, and Nobuhiro Go for a critical advice on the consistency principle.
This work was supported in part by a grant-in-aid
``platform for drug discovery, informatics, and structural life sciences'' from the MEXT, Japan.
\end{acknowledgments}
%\bibliography{refs,mypaper}
%\bibliographystyle{biophysics}

\end{document}